\documentclass[article]{aa}                             
\usepackage[varg]{txfonts}
\usepackage{graphicx}
\usepackage{array}
\usepackage{lscape}                                
\usepackage{txfonts}
\usepackage{caption}
\usepackage[normalem]{ulem}
\usepackage[dvipsnames]{xcolor}
\usepackage{booktabs}
\usepackage{siunitx}

\usepackage{longtable}
\usepackage{color}
\usepackage{supertabular}
\usepackage[draft]{hyperref}
\usepackage{siunitx}  
\usepackage[flushleft]{threeparttable}
\usepackage{multirow}
\usepackage{lscape}
\usepackage{placeins}

\newcommand{\Teff} {$T_{\rm eff}$}

\newcommand{\fastwind} {{\sc fastwind}}
\newcommand{\ioni}[2]{{#1\,\sc{#2}}}

\newcommand{\GG}{\mbox{$G$}}

\newcommand{\GGcor}{\mbox{$G_3^\prime$}}
\newcommand{\BPRP}{$G_{\rm BP,3}-G_{\rm RP,3}$}
\newcommand{\picor}{\mbox{$\varpi_{\rm c}$}}

\begin{document} 
%

\title{MONOS: Multiplicity Of Northern O-type Spectroscopic systems} 
\subtitle{III. New orbits and \textit{Gaia}--\textit{TESS} analysis for 10 SB2E systems}

\author{G.~Holgado\inst{1,2}, J. Maíz Apellániz\inst{3}, R.~C.~Gamen\inst{4,5}} 

\institute{Instituto de Astrof\'isica de Canarias, E-38200 La Laguna, Tenerife, Spain.
     \and
     Departamento de Astrof\'isica, Universidad de La Laguna, E-38205 La Laguna, Tenerife, Spain.
    \and
    Centro de Astrobiolog\'ia, ESAC campus, Camino bajo del castillo s/n, Villanueva de la Ca\~nada, E-28\,692, Spain
   \and
   Instituto de Astrof\'{\i}sica de La Plata, CONICET--UNLP, Paseo del Bosque s/n, La Plata, Argentina.
   \and
    Facultad de Ciencias Astron\'omicas y Geof\'{\i}sicas, Universidad Nacional de La Plata, Argentina.\\
}

\offprints{gholgado@cab.inta-csic.es}

\date{Date}

\titlerunning{MONOS III: \textit{Gaia}--\textit{TESS} new orbits}
\authorrunning{Holgado et al.}



  \abstract
{Understanding massive star multiplicity is essential for constraining formation scenarios, binary evolution, and stellar feedback in galaxies. The MONOS project (Multiplicity Of Northern O-type Spectroscopic systems) aims to characterize O-type spectroscopic binaries in the northern hemisphere ($\delta > -20^\circ$) combining high-resolution spectroscopy and multi-epoch photometry.}
{To use \textit{Gaia} DR3 and \textit{TESS} epoch photometry to identify orbital variability in O-type stars and derive orbital solutions.}
{We analyze ten O-type binary systems with high-quality \textit{Gaia} and \textit{TESS} photometry and available high-resolution spectra. We complement our analysis with additional photometric data from the \textit{Hipparcos} mission or the early-stage MUDEHaR survey. Periods are derived using three independent techniques and combined with radial velocity measurements to model each system using the \textsc{phoebe} code, yielding orbital and stellar parameters.}
{We present eight previously unpublished orbits---two with newly determined periods---and refine two others. In several cases, our periods match \textit{Gaia} values, though we highlight issues in automated determinations, such as half-period aliases. Among our notable discoveries, we point out the first known Oe+O spectroscopic binary (BD~$+$61~487) and a system of overcontact O-type supergiants in an eccentric orbit where significant mass transfer has taken place (HD~\num{169727}).
The derived solutions are consistent with spectral classifications and theoretical expectations, including short periods ($<3$ days), high mass ratios, and semi-detached or overcontact configurations.}
{These results expand the sample of O-type binaries with robust orbital characterization, especially in the short-period regime where tidal effects and mass transfer are prominent. The combined use of \textit{Gaia}, \textit{TESS}, and spectroscopy is shown to be effective, offering a scalable methodology applicable to southern hemisphere surveys. This work provides a foundation for future quantitative spectroscopic analyses and aids in achieving a comprehensive insight into the evolution of massive multiple systems.}


   \maketitle
%

\section{Introduction}

Massive O-type stars play a fundamental role in galactic evolution due to to their intense luminosities, powerful stellar winds, and eventual supernova explosions \citep{Bromm2009}. A striking feature of these stars is their high multiplicity fraction, with a large proportion found in binary or higher-order systems \citep{Masoetal98,Sana2012,Chini2012, Sotaetal14}.

Studying binarity in massive stars is crucial because interactions such as mass transfer, tidal synchronization, mergers, and overcontact configurations can dramatically influence their evolution and observable properties \citep{deMink2013, Sana2012, Sen2022}. Short-period O-type binaries are especially valuable for constraining key evolutionary processes like mass transfer and common-envelope evolution. These processes are needed to understand the formation of astrophysical objects such as X-ray binaries and gravitational wave progenitors.

Photometric variability due to eclipses, ellipsoidal modulation, pulsations, or magnetic activity offers powerful diagnostics of binarity and internal stellar physics \citep{Alfonso-Garzon2012, Southworth2022}. However, deriving accurate orbital parameters requires long-term, high-precision, and high-cadence photometric and spectroscopic datasets, which have only recently become widely available thanks to missions such as \textit{Gaia} and \textit{TESS}.

The MONOS project (Multiplicity Of Northern O-type Spectroscopic systems; \citealt{MaizApellaniz2019,Trigetal21}) is systematically characterizing northern ($\delta > -20º$) O-type binaries, complementing southern hemisphere efforts like the OWN survey and the MOSOS project. Large compilations of massive star orbits, such as those by \citet{Sana2012}, \citet{Chini2012}, \citet{Mason2009}, and \citet{Malkov2020}, have significantly advanced our understanding of multiplicity and the overall population of massive binaries. Nevertheless, detailed orbital solutions remain scarce relative to the total number of known massive binaries, particularly for systems without previous spectroscopic follow-up or with incomplete or poor orbital determinations. 
Our work, adding eight new orbital solutions, represents a significant step forward in improving the census of well-characterized massive star binaries.

In this work, we focus on a subsample of short-period O-type binaries, chosen both for their crucial role in constraining mass transfer and common-envelope evolution, and for the practical advantage that their brief orbital periods allow radial velocity measurements to be obtained over relatively short timescales. This makes them particularly suitable for timely orbital characterization. Our study thus constitutes a first step toward a more complete analysis of the entire MONOS sample, which includes longer-period systems that require extended observational monitoring over several years. We combine multi-epoch photometry from \textit{Gaia} and \textit{TESS} with high-resolution spectroscopy to derive new or refined orbital solutions for this subsample. We present orbital analyses for ten systems: eight new orbits and two benchmark cases from the literature. The structure is as follows: Sect.\ref{obser} describes the observational data, Sect.\ref{methods} outlines the analysis methodology, Sect.\ref{Results} discusses individual system results, and Sect.\ref{conclusions} summarizes our conclusions.

\section{Observational data}\label{obser}

\subsection{Epoch photometry}

Our study is based on the data that \textit{Gaia} has provided in the latest Data Release, the epoch photometry of \textit{Gaia} DR3 \citep[\textit{Gaia} Collaboration Vallenari][and {\url{https://doi.org/10.17876/gaia/ dr.3/55}}]{Vallenari2023}.
The total number of variable sources published in DR3 is 1\,811\,709\,771, with 11\,754\,237 sources having epoch photometry time series.
The characteristics of these photometric observations are compiled in \cite{Eyer2022}. The main ones are that the observations encompass a period of 34 months (25 July 2014 to 28 May 2017) with a slightly higher cadence for objects near the ecliptic poles and  ecliptic latitude $\beta$ = $\pm$45. 
The average number of points available per system is $\sim$50 epochs.
The measured variations are on average 0.001 mag for a range of magnitudes of \GG\ between 3.5 and 23 mag \citep{Evans2022}.

For the \textit{TESS} data, we constructed light curves using the open-source package \textsc{tglc}, which generates high-precision photometry by modeling the effective point-spread function (PSF) of each source to reduce contamination from nearby stars \citep[see][]{2023AJ....165...71H}. This method incorporates stellar positions and magnitudes from \textit{Gaia} DR3 as priors. The resulting \textit{TESS}-\textit{Gaia} light curves (TGLCs) reach a photometric precision consistent with the mission’s prelaunch noise-level predictions.
Each system in our sample is observed, on average, in three \textit{TESS} sectors (i.e., $\sim$80--100 epochs per sector, or $\sim$250--300 data points in total), although in two cases only a single sector is available. The \textsc{tglc} pipeline provides both PSF and aperture photometry. After comparison, we adopted the aperture fluxes, as they showed better internal consistency across sectors for our targets.
Due to \textit{TESS}’s large pixel size ($\sim$21 arcsec/px.), aperture photometry is especially susceptible to flux contamination in crowded fields, requiring individual inspection of nearby sources despite mitigation efforts.
The aperture fluxes were converted to magnitudes following the standard \textit{TESS} calibration\footnote{\url{https://tess.mit.edu/public/tesstransients/pages/readme.html}}. Because our analysis involves comparison with magnitudes from other photometric bands, we incorporated an additional systematic uncertainty to account for the zero-point offset. Following the recommendations in the \textit{TESS} handbook, we adopted a conservative zero-point uncertainty of 0.05~mag, which dominates the total error budget in our magnitude-based comparisons. This is reflected in the error bars shown in Sect.~\ref{Results}.

Although both datasets are used in parallel, greater weight is given to the \textit{Gaia} epoch photometry, especially in cases of discrepancy (e.g., DN~Cas in this work), due to its higher spatial resolution and lower susceptibility to contamination. Nonetheless, the \textit{TESS} light curves play a key role in constraining the eclipse morphology, providing essential information on the orbital inclination and relative stellar radii thanks to their high cadence and continuous coverage.

We have also used the photometric results from the MUDEHaR project. These are characterized by precise H$\alpha$ and Ca\,{\sc ii} window variability measures, through two narrow filters, on the scale of days-months-years, for tens of thousands of stars (from AB mag = 3 down to 17 with S/N between 70 and 500) \citep[see][and Holgado et al. in prep.]{Holgado2023,Holgado2024}. The small pixel size (0.55 arcsec/pix) naturally overcomes the crowding and nebular problems that affect other surveys such as \textit{Gaia} or \textit{TESS}. The acquisitions are derived from four distinct exposure durations (0.1+1+10+50 seconds). Each source is planned to be visited a maximum of 20 times throughout the year, capturing a minimum of 5 epochs on each occasion. This enables the assessment of variability on an hourly scale. The final result would be 100 epochs per year for each field (i.e. each star). 
In this study, we found MUDEHaR data available for one of our targets. However, we anticipate that MUDEHaR could play an increasingly important complementary role in future studies, especially in cases where the primary photometric datasets are unavailable or limited. 

\subsection{Spectroscopy}

Spectroscopic data were obtained from LiLiMaRlin (Library of Libraries of Massive-Star High-Resolution), a project that compiles high-resolution spectroscopy from various spectrographs across both hemispheres, ensuring a consistent data reduction process. This includes normalization, telluric-line removal, defect and problematic spectrum elimination, and uniformization \citep{Maizetal19a}.
The collected data originate from dedicated projects such as the
ALS catalog \citep{Pantetal21},
OWN survey \citep{Barba2017},
IACOB project \citep{SimonDiaz2015,SimonDiaz2020},
CAFE-BEANS \citep{Neguetal15a}, and
NoMaDS \citep{Pelletal12}, along with archival searches in public databases. The spectra used for the sample in this paper were obtained with (a) HERMES at the 1.2~m Mercator Telescope, FIES the 2.6~m Nordic Optical Telescope, and HARPS-N at the 3.6~m Telescopio Nazionale Galileo at the Observatorio del Roque de los Muchachos (ORM), La Palma, Spain; (b) FEROS at the MPG/ESO 2.2~m Telescope at the Observatorio~de~La~Silla, Chile; and (c) HRS at the 10~m Hobby-Eberly Telescope at the McDonald Observatory, USA.

\subsection{Additional data}

The main results of this paper are presented in Table~\ref{PHOEBE_params} for the eight stars without previous spectroscopic orbits and in Tables~\ref{HD1810A_table}~and~\ref{PHOEBE_DNCas} for the other two. In those tables we also show reference information for each star from \textit{Gaia}~DR3. More specifically, we give the \GGcor\ magnitude and the \BPRP\ color, with the first one including the calibration corrections of \citet{MaizWeil24}. We also give the corrected parallax applying the calibration of \citet{Maizetal21c,Maiz22} and the corresponding \textit{Gaia}~DR3 distance from the ALS~III paper (Pantaleoni González et al. submitted to MNRAS), which uses the Bayesian model of \citet{Maiz01a,Maiz05c} with the updated parameters from \citet{Maizetal08a}, which are the most appropriate approach for the spatial distribution of Galactic OB stars.

\section{Methodology}\label{methods}
The primary goal of this paper is to study the information on the variability of Galactic O stars contained in \textit{Gaia} DR3 epoch photometry with \textit{TESS} information available in order to obtain orbits in combination with the spectra available. In this section we describe how we selected the sample, compiled the data, calculated the periods and light curves, and classified the results.

\subsection{Photometric variability and sample selection}

We searched the \textit{Gaia}~DR3 archive for Galactic O-type stars in ALS with epoch photometry and we found 154 sources. The range of magnitudes in \textit{Gaia}˜DR3 \GG\ for this sample is between 3.9 and 14.0 mag, the majority ($\sim$80\%) between 6.2 and 11.4 mag. Our first goal was to determine for which stars in the sample we can obtain a period and, for those, obtain a light curve and establish its orbit.
The average number of epochs for each \textit{Gaia} source is 50, and 80\% of the sources have between 25 and 60 epochs. There is a correlation between the number of epochs and the capability to achieve a robust period analysis, with only 25\% of sources having fewer than 25 epochs providing a reliable period result. 

The 154 sources selected are those with \textit{Gaia} epoch photometry available and were cross-matched with entries in the \textit{TESS} database containing available photometric information.
Using the combined photometric data from \textit{Gaia} and \textit{TESS}, we were able to derive periods (see below) for 70 of these systems.
The vast majority (85\%) exhibited short periods, under five days.
At this stage, we refined the sample to identify the most relevant candidates for the specific goals of the MONOS project. This selection focused on objects observable from the Northern Hemisphere (32 in total), for which we obtained robust and stable periods, and that met at least one of the following criteria: either no confirmed period was available beyond the automatic value derived from the \textit{Gaia} analysis, or, although a period was reported in the literature, no published orbital solution for the system existed.
Following this selection process, eight sources remained and are analyzed in this work.
Additionally, two well-studied stars from the literature were included as benchmarks to assess the robustness of our methodology, bringing the total number of systems analyzed in this study to ten.
For the selected cases, light curves were generated using the \textsc{tglc} Python script (see Sect.~\ref{obser}).
We visually inspected the resulting light curves and compared them with those from \textit{Gaia}. Based on this analysis, we chose to use aperture photometry, which exhibited fewer artifacts and yielded more stable results.
For the 10 selected O-type star systems in this study, we applied three independent period determination methods, ensuring consistency and to reinforce the robustness of our approach. These methods were initially tested on the larger general sample of 154 systems, finding a perfect fit for 70 of them, where they showed excellent agreement. The broader comparison with the literature has been highly satisfactory, further validating our results.

We used periodograms \citep{Lomb1976,Scargle1982}, phase dispersion minimization (PDM) \citep{Stellingwerf1978,Cincotta1995}, and the String Minimization Method (a two-dimensional generalization of the previous PDM). Each analyzes light curves differently but consistently identified stable periods. The agreement across methods confirms the reliability of our determinations.

Only the periodogram method provides a formal uncertainty, often too small. To improve accuracy, we compared period deviations across the three \textit{Gaia} filters and manual adjustments. In 53\% of cases, the filter-based uncertainty exceeded the manual one, confirming reliability. The larger value was adopted as the final uncertainty for the period.

\subsection{Spectroscopic analysis}

Spectroscopic data are needed to determine the sepctroscopic status, measure the radial velocities to determine the spectroscopic orbit, and derive the spectral classifications of the observed components.

We determine, for each system, the current spectroscopic status (classification + eclipsing multiplicity nomenclature) as described in the first MONOS paper \citep{MaizApellaniz2019}. Our goal is to progressively include such a nomenclature for the full ALS sample as new data are gathered and it includes the information obtained on this paper regarding eclipsing binaries and/or ellipsoidal variables (denoted as E in both cases).

For 9 out of the 10 systems in our sample we have between 8 and 15 LiLiMaRlin epochs (many of them from the IACOB project). Specifically, for five systems the spectra came from a dedicated observational campaign designed to cover the orbits of these short-period O-type binaries. This campaign ensured systematic coverage of most orbital phases, and it was awarded time, emphasizing the importance of obtaining detailed orbital analyses for these systems. For the remaining system (GLS~\num{19307}), the faintest one in the sample, we only have two LiLiMaRlin epochs, which only allowed us to do a less detailed analysis, as described below. The procedure used to analyze the stars was:

\begin{itemize}
  \item A preliminary determination of the radial velocities (RV) was done using \textsc{iraf}, typically analyzing the \ion{He}{i} $\lambda4542$, $\lambda4922$, $\lambda5411$ or $\lambda5875$ lines. We used \textsc{splot} and, in poorly separated double lined systems, the \textsc{ngauss} task. 
  \item The obtained RVs were then used to determine a preliminary spectroscopic orbit  through the \textsc{gbart} code \citep{2017ascl.soft10014B}
  \item We then used the spectral disentangling code \textsc{unwind} (Maíz Apellániz et al. in prep.) and the preliminary orbit to (a) measure the ISM spectrum of each sightline (to subtract it from each epoch), (b) disentangle the two components  (three in the case of BD~$-$19~4929), (c) measure the RVs again using the disentangled profiles, and (d) determine a new spectroscopic orbit. The process was iterated as needed (typically twice was enough).  
  \item As a final step for the spectroscopic orbit determination, we generated a \fastwind\ model for each component and cross-correlated it with the \textsc{unwind} output for the purpose of refining the systemic velocity $\gamma$ value. This step is required because O stars have different degrees of wind infilling in each line that affect the effective central wavelengths. The accuracy of the $\gamma$ values is of no consequence for e.g. the mass determinations but is critical for determining whether the star is a walkaway or runaway \citep{Maizetal18b}.
  \item The resulting disentangled spectroscopic components were degraded to the spectral classification resolution of 2500 and compared to a standard spectral library derived from the Galactic O-Star Spectroscopic Survey (GOSSS, \citealt{Maizetal11}) and Alma Luminous Star (ALS, \citealt{Pantetal21}) projects using the \textsc{mgb} code \citep{Maizetal12} to obtain their spectral classifications \citep{Maizetal24c}.
\end{itemize}

\subsection{Spectro-photometric study}\label{CirculMethod}

We have employed the legacy version of \textsc{phoebe} \citep[PHysics Of Eclipsing BinariEs][]{Prvsa2016}, an open-source package for modeling eclipsing binaries, with the aim of determining orbital and stellar parameters, and an approximated eccentricity value, and inclination, when no radial velocities are available.

For each system, we have employed \textit{Gaia} and \textit{TESS}\footnote{and MUDEHaR in one case} epoch photometry data, the calculated period, and spectral type information from ALS to allocate initial temperatures and masses to the system's components. Additionally, \textsc{phoebe} demands various inputs such as the model type to fit and the gravity and limb-darkening characteristics of the system.
Following several iterations, we have determined values that yield the optimal fit for the light and radial velocity curves.
In the one instance where the radial velocity data was insufficient, the iterative fitting procedure exerted a negligible influence on the mass ratio and semi-major axis parameters employed in \textsc{phoebe}.

In every instance, we accounted for the plausible third light to address the contamination stemming from alternative sources, with particular emphasis on the \textit{TESS} data, as well as in other systems identified as triple or multiple within the spectroscopic data. A detailed discussion of each case is below.

For each system, we also calculated the fillout factor, which quantifies the degree of Roche lobe filling by the stellar components. This was done following the formalism presented in \citet{Mochnacki1972}, which compares the stellar surface potential to the critical Roche lobe potential. This dimensionless quantity allows us to quantify how close a star is to filling (or overflowing) its Roche lobe, with values approaching unity indicating systems in contact or near-contact configurations. The fillout factor thus provides a useful diagnostic for assessing the evolutionary status of the binaries, especially in the context of possible mass transfer or overcontact phases.
The computed fillout factors are included in Table~\ref{PHOEBE_params}, and are discussed in the context of individual systems in Section~\ref{Results}.

\section{Individual systems}\label{Results}

\subsection{Systems with new spectroscopic orbits}

\subsubsection{BD~$+$61~487 = GLS~\num[detect-all]{7485}}


\begin{figure}
    \centering
    \includegraphics[width=0.49\textwidth]{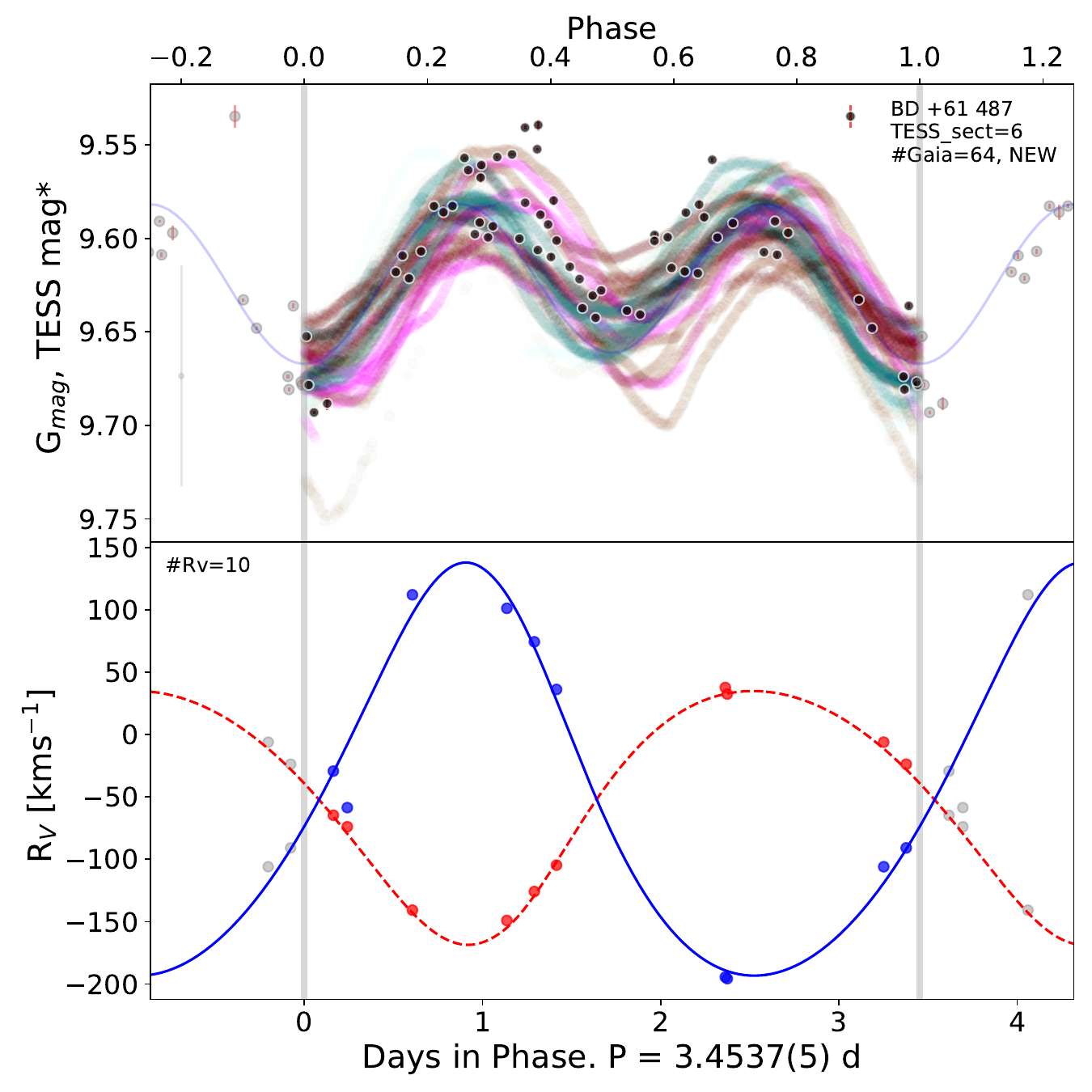}
    \caption{(top) Folded light curve of \textit{Gaia} G magnitude data and \textit{TESS} available sectors for BD~$+$61~487.
In black \textit{Gaia} epoch-photometry and in light colors the different sectors of \textit{TESS}. 
\textit{Gaia} Observations (black dots) include uncertainties in red. 
The grey point with uncertainty in the corner marks the typical error for \textit{TESS} magnitudes considering the zero point uncertainty.
The orbital \textsc{phoebe} model obtained for the set of points is included in light blue.
In grey are repeated points beyond the central period. 
The number of \textit{Gaia} points, \textit{TESS} available sectors, and status are indicated in the figure.
(bottom) spectroscopic orbit with radial velocity measurements for the primary and secondary, and the number of spectra used. Lines coming from \textsc{phoebe} model.}
    \label{BD_+61_487_3p4537(5)_column}
\end{figure}

Using GOSSS spectra, eight years ago we discovered (but did not get to publish) that this SB2E system was an Oe+O system. Oe stars are relatively uncommon and none were known to have a type-O spectroscopic companion until now, so this discovery may prove to be an important clue towards understanding their formation. Since then, another example of an Oe+O system has been discovered in GOSSS data, HD~\num{156172}, and that object will be discussed in a future paper.

\begin{figure}
    \centering
    \includegraphics[width=0.45\textwidth]{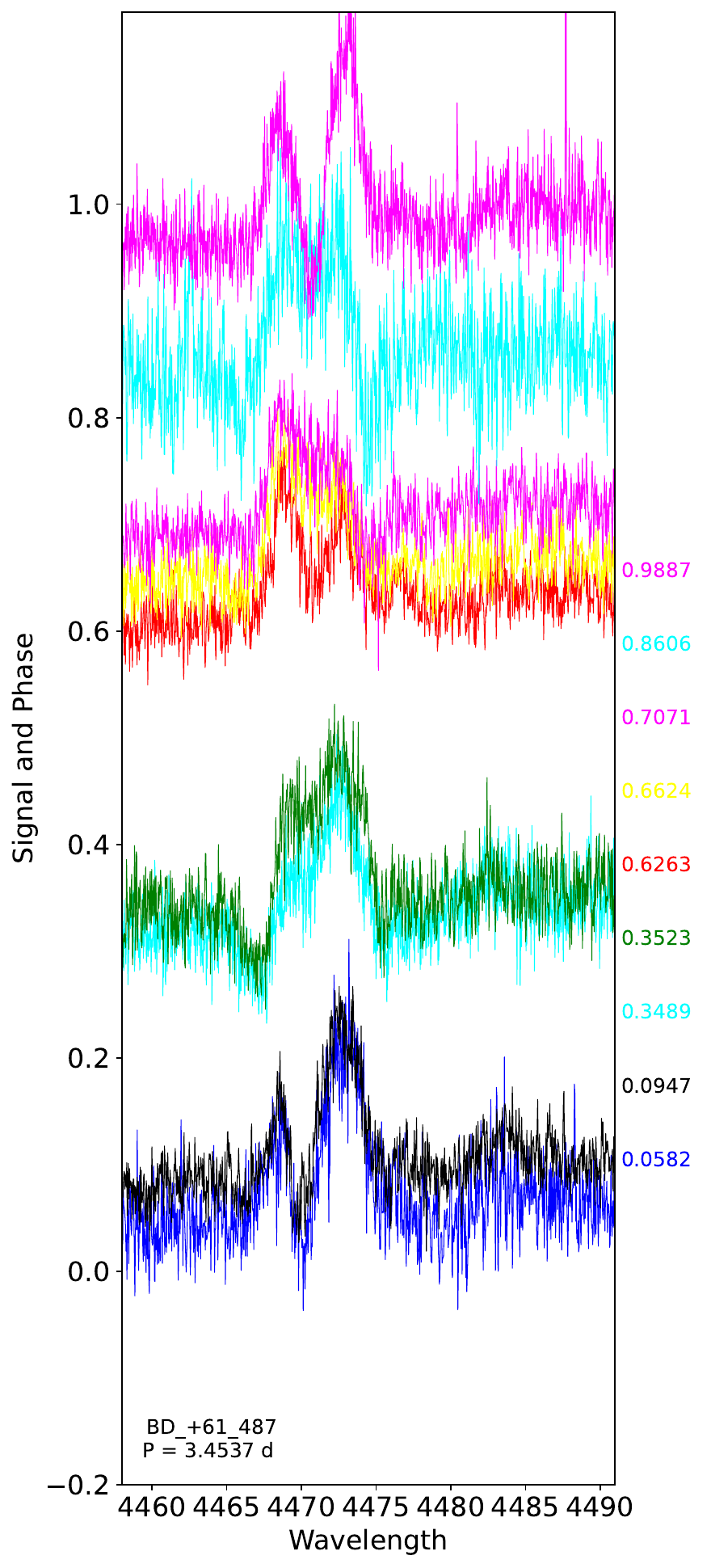}
    \caption{The composite spectrum of BD~$+$61~487 adjusted in height to the phase of each of the observations, for the \ion{He}{I} $\lambda$~4471 line. The color is assigned randomly, and the orbital phase of each spectrum—matching the same color—is displayed on the right side. This figure in combination with Fig.~\ref{BD+61487_4541} (in Sect.~\ref{Spectra}) show the Oe+O nature of the system.}
    \label{BD+61487_4471_in.pdf}
\end{figure}

BD~$+$61~487 has most lines in emission, with the most notorious exceptions being \ioni{He}{ii} lines. The existence of strong emission from the Oe component in several lines (including \ioni{He}{i}~$\lambda$4471, see Fig.~\ref{BD+61487_4471_in.pdf} and Figs in Sect.~\ref{Spectra}) precludes the determination of subtypes but the existence of two absorption lines in \ioni{He}{ii}~$\lambda$4542 that move with the orbit identifies the two components as O stars. We obtain a period of \num{3.4537}(5)~d, which is, to our knowledge, the first determination of a photometric period for the system. The primary and secondary minima are significantly different and the orbit appears to be non-circular but the light curve is dirtier than for most ellipsoidal variables, likely a sign of the additional variability introduced by the Oe component.

The observed spectra (Sect.~\ref{Spectra}) clearly reveal spectral features from three distinct components: a pair of absorption lines with markedly different widths and double-peaked emission lines from a disk. Analyzing the spectra in the context of the photometric periodicity, we identified \ion{He}{ii} $\lambda$4200, 4542, and 5412 absorption lines characteristic of an SB2 system, along with \ion{He}{ii} $\lambda$4686 emission, which follows the motion of the narrow absorption lines. The disk emissions also track the narrow-lined component but with a significantly lower semi-amplitude than the absorptions. The broad-lined component appears to exhibit less motion than the narrow one, suggesting it is the more massive star. There are no signs of the \ion{O}{III} $\lambda$5992 line, which would indicate a relatively late spectral type for the narrow component. In the broader component, this line could be absent due to dilution. 

We measured \ion{He}{ii} $\lambda$4542 and 5412 using \textsc{ngauss@iraf} to obtain a preliminary RV orbital solution, processed with \textsc{unwind} to improve it, and then combined with photometric data to develop a \textsc{phoebe} model of the system. In the spectra disentangled with \textsc{unwind} there are strong residuals. This is likely caused by the typical line variability in Oe stars. In the future we plan to observe the system several times during two consecutive orbits to minimize the effect of the temporal line variability and better constrain the spectral classification and reduced the uncertainty in the (significant for a short-period system) eccentricity. In the meantime, we can only shown a non-disentangled spectra in Fig.~\ref{spectra3} and provide an Oe+O classification. In this case we could not generate a \fastwind\ model, so the values of the systemic velocity $\gamma$ in Table~\ref{PHOEBE_params} do not include the additional correction.

In Fig.~\ref{BD_+61_487_3p4537(5)_column}, we present the results of the \textsc{phoebe} model fit (parameters in Table~\ref{PHOEBE_params}). No previous orbital solutions were found in the literature. The photometric observations are sourced from \textit{Gaia} and \textit{TESS} in sectors 18, 19, 58, 59, 85, and 86, with a specific shift applied to \textit{TESS} data to align it with the \textit{Gaia} measurements. The model does not fit precisely in both cases, likely due to the disk influencing the measurements.
We remark that, for the first time, this work determines and presents the orbital parameters and period of the system, along with the associated light and radial velocity curves.


\subsubsection{HDE~\num[detect-all]{255312} = GLS~\num[detect-all]{8891}}


\begin{figure}
    \centering
    \includegraphics[width=0.49\textwidth]{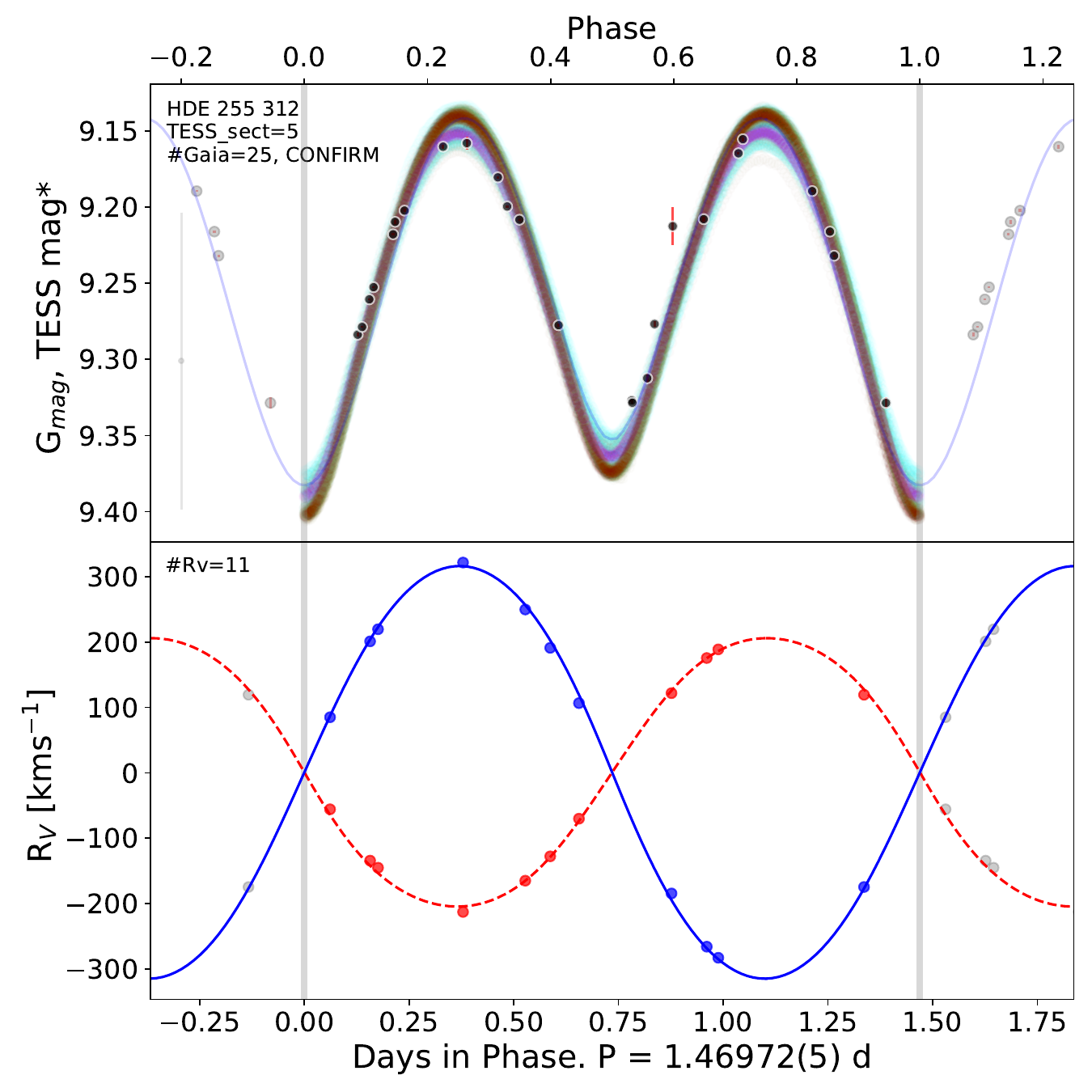}
    \caption{Same as Fig.~\ref{BD_+61_487_3p4537(5)_column}, for the HDE~\num{255312}.}
    \label{HDE_255_312_1p46972(5)_column}
\end{figure}

This system was first identified as an O+O binary by \citet{Li21}. Our calculated period is \num{1.46972}(5)~d, representing relatively novel information, as the only previous identification comes from \citet{Mowlavi2022}, where a similar period was estimated using automated methods without visual representation of the light curve, and no radial velocity confirmation.

\begin{figure*}
    \centering
    \includegraphics[width=\textwidth]{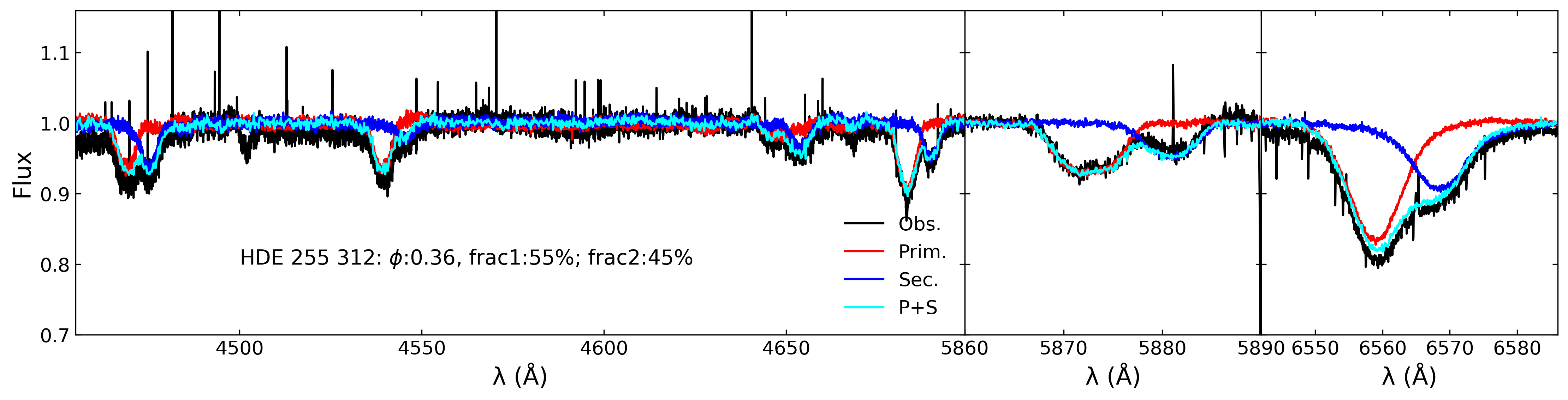}
    \includegraphics[width=\textwidth]{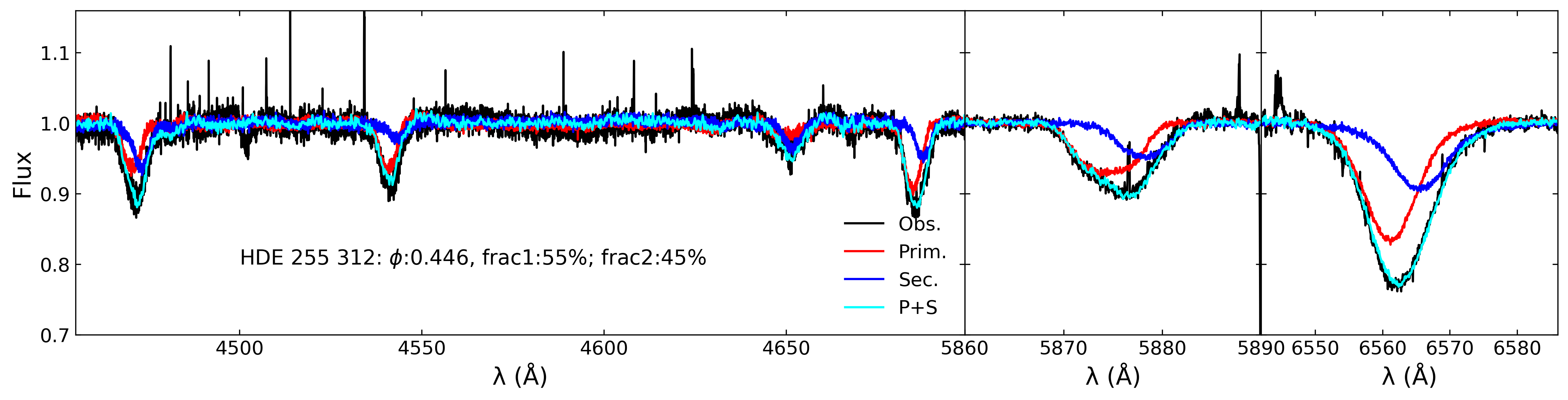}
    \caption{Examples of spectral disentangling for HDE~255~312 at orbital phases $\phi$ = 0.36 and 0.446. The observed spectrum (black) is shown together with the disentangled spectra of the primary (red) and secondary (blue) components. The synthetic composite spectrum (cyan), obtained by summing the disentangled components weighted by their flux contributions (frac1: 55\%, frac2: 45\%), is also displayed. The agreement between the observed and reconstructed spectra demonstrates the quality of the disentangling performed with the UNWIND program.}
    \label{HDE255312_1}
\end{figure*}

The spectra exhibit two clear and distinct components in all \ioni{He}{i} and \ioni{He}{ii} lines with the anticipated oscillatory motion present (see Fig.~\ref{HDE255312_1}). The \ioni{He}{i}~$\lambda$4922 lines were employed to derive radial velocity measurements and obtain a preliminary orbit.
From that, we applied \textsc{unwind} to obtain the disentangled spectra and the final spectroscopic orbit, which was then combined with photometric data to fit a \textsc{phoebe} model. 

The disentangled spectra yield a classification of O7 V((f))(n)z~+~O9.5 V(n), with spectral subtypes somewhat later than those of \citet{Li21}. The weakness of \ioni{O}{iii}~$\lambda$5592 is atypical for the assigned spectral types and may be associated with relatively high rotational velocities and dilution effects.  

Fig.~\ref{HDE_255_312_1p46972(5)_column} illustrates the comparison between the \textsc{phoebe} model fit and the observational data. No prior orbital representations have been identified in the existing literature. The photometric observations are derived from \textit{Gaia} as well as \textit{TESS} sectors 43, 44, 45, 71, and 72. With this number of spectra (n=11), the fit is remarkably accurate, and the orbit determination appears robust. The parameters derived from the \textsc{phoebe} model fit are enumerated in Table \ref{PHOEBE_params} and align with the theoretical expectations for the given spectral types. 
In the study by \cite{Cruzalebes2019}, a radius measurement was acquired via mid-infrared interferometry; nonetheless, both their assigned spectral type (B2) and the radius value of 67.2 R$_{\odot}$ are inconsistent with our results.
Finally, we note that the best-fit \textsc{phoebe} model corresponds to an overcontact configuration, a conclusion further supported by the fact that the calculated fillout factor exceed 1. Overcontact binaries represent a rare and astrophysically significant class of systems, and the discovery of each new case is valuable \citep{Abdul-Masih2025}.


\subsubsection{HD~\num[detect-all]{169727} = GLS~\num[detect-all]{5046}}


\begin{figure}
    \centering
    \includegraphics[width=0.49\textwidth]{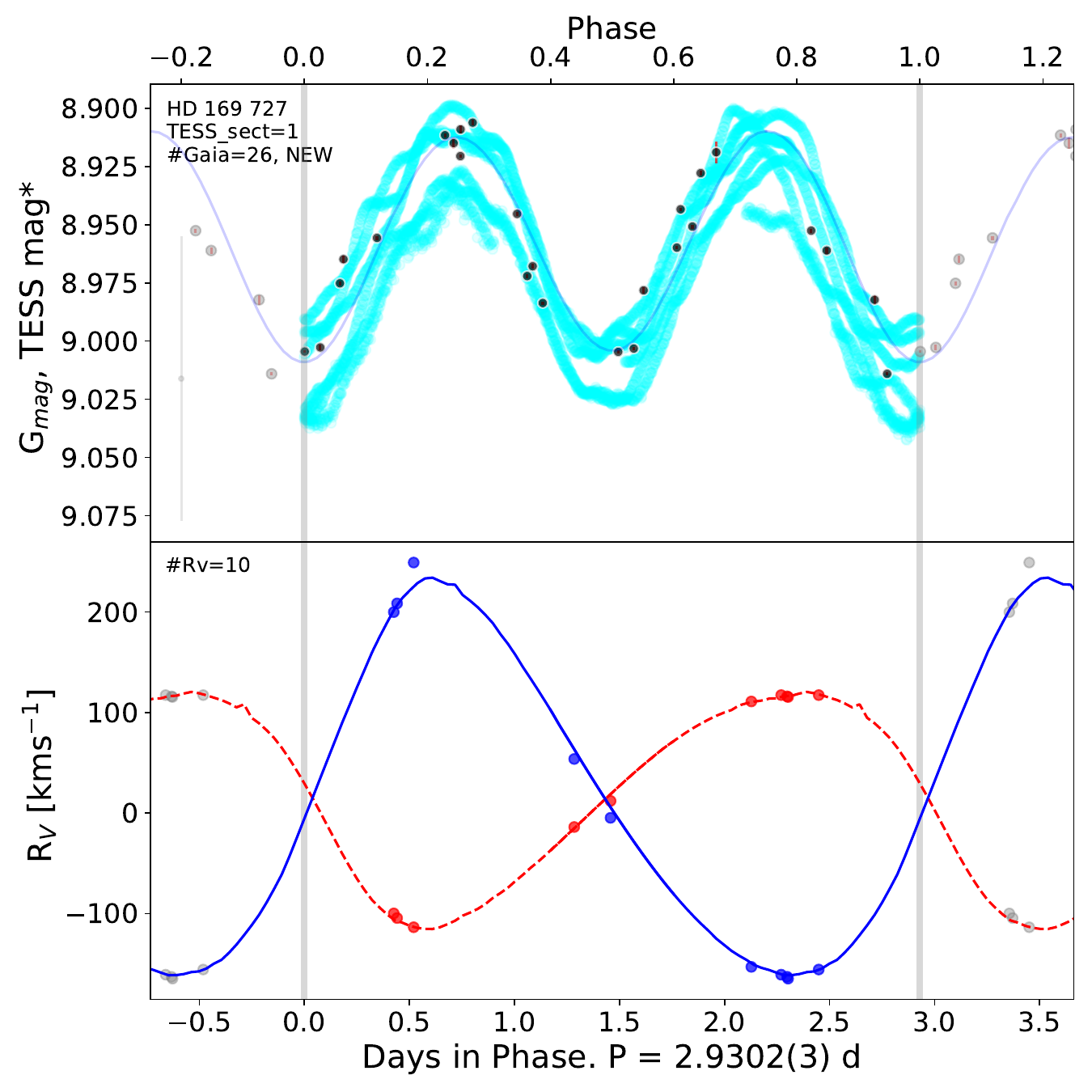}
    \caption{Same as Fig.~\ref{BD_+61_487_3p4537(5)_column}, for the HD~\num{169727}.}
    \label{HD_169_727_2p9302(3)_column}
\end{figure}

HD~\num{169727} has received spectral classifications as an O star since the 1950s \citep[see, e.g.,][]{Roman1955,Vijapurkar1993} but the literature includes notorious outliers, some as late as B8 \citep{Houk1988}, reflecting the disparity in the quality of spectral classifications. To our knowledge, nobody has reported it as an O+O SB2E, which we do here. Our calculated photometric period is \num{2.9302}(3)~d and, also to our knowledge, it is the first instance in the literature that a period is derived for the system.
\citet{Mowlavi2022} does not provide a value for the period and \citet{Rimoldini2022} classifies it as an RS. 
 
Our available spectra for this system shows separated \ioni{He}{i} and \ioni{He}{ii} lines with a narrow component and a broader component (Figs.~\ref{HD169727_1}). We used the \ioni{He}{i}~$\lambda$4922 and 5875 lines to obtain the radial velocity measurements and derive a preliminary orbit, followed by \textsc{uniwnd} runs to disentangle the spectra and obtain an improved orbit. We obtain spectral types of O7.5~Iabfp and O7~Iabf from the disentangled spectra. Having two supergiants in such a short-period orbit already points out towards the possibility of an interaction. Another further indication is that the A component (defined as the most massive, see below) is an Onfp star \citep{Sotaetal11a}, that is, an Of star with \ioni{He}{ii}~$\lambda$4686 emission with an absorption reversal, though in this case missing the n characteristic (possibly because of the low inclination, see Table~\ref{PHOEBE_params}). The two photometric minima are quite similar, pointing towards ellipsoidal variability rather than eclipses. Again, the weakness of \ioni{O}{iii}~$\lambda$5592 is linked to high rotational speeds and dilution.

Figure \ref{HD_169_727_2p9302(3)_column} shows the \textsc{phoebe} model along the observational data used to calibrate it.  Photometric data are sourced from both \textit{Gaia} and \textit{TESS} sector 80. The \textsc{phoebe} parameters are listed in Table~\ref{PHOEBE_params}, including a large eccentricity value of $e \sim 0.15$, and an overcontact configuration for the fitted model. 
The optimal \textsc{phoebe} model confirms an overcontact configuration, supported by a fillout factor greater than 1.

However, the most peculiar of our results is what we find for the masses. By definition, component A is the more massive star, and the mass ratio indicates that it is significantly more massive than B. However, the luminosity ratio in the $B$ band is inverted, with the secondary being more than one magnitude brighter (i.e., a luminosity ratio of approximately 1/3). Given that the radii of both components are similar while their masses differ substantially, the system's configuration is not straightforward to interpret.
We interpret these features as indicative of an advanced evolutionary interaction: the current secondary component likely evolved first, becoming brighter and subsequently transferring a substantial fraction of its mass to the current primary, which has not yet had time to evolve significantly. 
The notably high eccentricity is unexpected for an overcontact system, although the relatively long orbital period, exceeding 2.9 days, could contribute to preserving such an orbital characteristic.
In the literature, it appears in a list of massive X-ray-detected stars \citep{Naze2009}, which may be related to its overcontact configuration.  


\subsubsection{BD~$-$13~4929 = GLS~\num[detect-all]{4913}}

\begin{figure}
    \centering
    \includegraphics[width=0.49\textwidth]{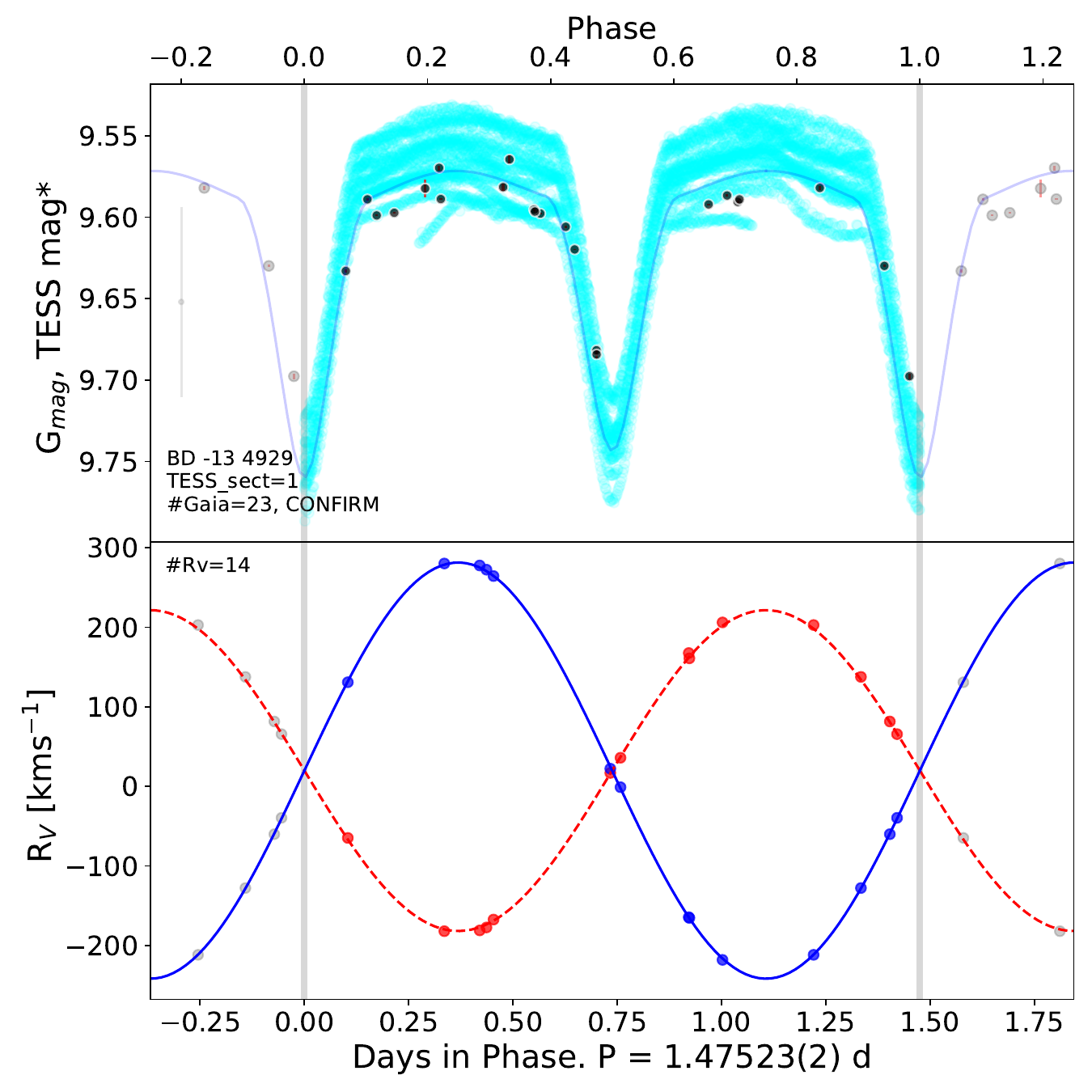}
    \caption{Same as Fig.~\ref{BD_+61_487_3p4537(5)_column}, for the BD~$-$13~4929.}
    \label{BD_-13_4929_1p47523(2)_column.pdf}
\end{figure}

This system was identified as an SB3 by \citet{Sana2009}. Our calculated photometric period is \num{1.47523}(2)~d (similar to the one given by \citealt{Mowlavi2022}). 
\citet{Maizetal22a} reported spectral types of O8~V~+~B0.5:~V~+~B0.5:~V, with our new classification based on disentangled spectra yielding similar results: O8~V~+~B0.2~V~+~B0.5~V.
The two early-B stars are the eclipsing binary, so the type is SB2E+C. The \textit{Gaia}~DR3 distance listed in Table~\ref{PHOEBE_params} is the one for the star but note that, as BD~$-$13~4929 is in M16 (= Villafranca~O-019), one can also use the more precise distance for the cluster from \citet{Maizetal22a} of $1234\pm 16$~pc. The difference in terms of the uncertainty of the stellar distance is slightly over 2 sigmas, at the edge of the range allowed if the star is really located in M16. The improvement that \textit{Gaia}~DR4 will bring to the astrometry should decide this issue, most likely by increasing the stellar parallax\footnote{If instead of using the calibration of \citet{Maizetal21c,Maiz22} one assumes that the published random uncertainty of the \textit{Gaia}~DR3 parallax is correct, then the difference would be $\sim$ 4 sigmas, putting BD~$-$13~4929 clearly beyond M16. This is a good example of the need for an accurate calibration of \textit{Gaia} parallaxes.}.

The spectra distinctly show the orbital dynamics of the B-type star pair in the \ioni{He}{i} lines, while the O-type star remains stationary (Fig.~\ref{BD-134929_1s}). In contrast, the influence of the B-type stars is barely perceptible in the \ioni{He}{ii} lines, just marginally in \ioni{He}{ii}~$\lambda$4686. 
The \ioni{O}{iii}~$\lambda$5592 line is visible and exclusively associated to the O-type star. The line is relatively narrow, indicating a moderate rotation.
Furthermore, a very faint companion component appears towards the red wing of this line, motion-less for the entire cycle, thus not related with the B-type stars detected in the \ioni{He}{i} lines. 
This observation might suggest the existence of a very hot yet faint companion to the O-type star, although its stationary character presents interpretative challenges.

The \ioni{He}{i}~$\lambda$4922 lines were employed to determine a preliminary orbit, the data were then further processed by \textsc{unwind}, and the results were subsequently integrated with photometric data to produce and calibrate the \textsc{phoebe} model. \textsc{unwind} requires an extra iteration due to the presence of a third component but is able to provide a very clear disentangling of the system, yielding spectral classifications of O8~V~+~B0.2~V~+~B0.5~V, which are consistent with the \citet{Maizetal22a} result (which did not use disentangling and, therefore, had some uncertainty in the B subtypes). The most massive of the two B stars is the earliest one, as expected (but see below for the values themselves).

Figure~\ref{BD_-13_4929_1p47523(2)_column.pdf} displays the alignment of the light curve and spectroscopic orbit of this eclipsing star with our best \textsc{phoebe} model. 
The corresponding \textsc{phoebe} parameters are listed in Table~~\ref{PHOEBE_params}. 
Although the research conducted by \citet{Sana2009} did not provide an orbital solution, they proposed a mass ratio \(q=1\) and an approximate orbital period of $\sim$4 days for the B-type star pair. 
Their study also outline the hierarchical system we see now clearly in our spectra. 
Calculating the influence of third light was essential for this \textsc{phoebe} fit. We identified the O-type star as this third light, accounting for 64\% of the total flux of the system. 
This estimate was derived from the spectral types considered here, and its absolute magnitudes according to the \cite{Martins2005} calibration for the O-type stars. 
For the B-type stars we use an approximation given the the magnitudes of the stars reported in \cite{Nieva2014}.
On the whole, despite the limited number of \textit{Gaia} data points and having only one \textit{TESS} sector (80) available, the orbit appears robust.
The radial velocity data obtained from the spectra align closely with the predictions by the \textsc{phoebe} model (see also Fig.~\ref{BD-134929_1s}).

The parameters derived from the \textsc{phoebe} model are presented in Table~\ref{PHOEBE_params}, with both the masses and radii below the typical values for B0.2-0.5~V stars, possibly indicating some degree of stripping. 
This scenario of partial envelope stripping has been investigated in theoretical models \citep[e.g.,][]{Gotberg2017} and is supported by observational evidence from confirmed stripped stars in binary systems \citep{Ramachandran2023, Ramachandran2024}.

The effective temperature we derive aligns well with the values reported by \cite{Wolff2006} and those listed in the PASTEL catalog \citep{Soubiran2016}. However, the derived mass shows a significant discrepancy with respect to \cite{Wolff2006}, likely due to their analysis not accounting for the separation of the luminosity contribution from the O-type component. In contrast, \cite{Dufton2006} specifically focus on the O-type component, which they classify as an O9 star. It is therefore essential to avoid a direct comparison between their results and ours, which refer exclusively to the B-type components. 



\subsubsection{BD~$+$60~498 = GLS~\num[detect-all]{7269}}


\begin{figure}
    \centering
    \includegraphics[width=0.49\textwidth]{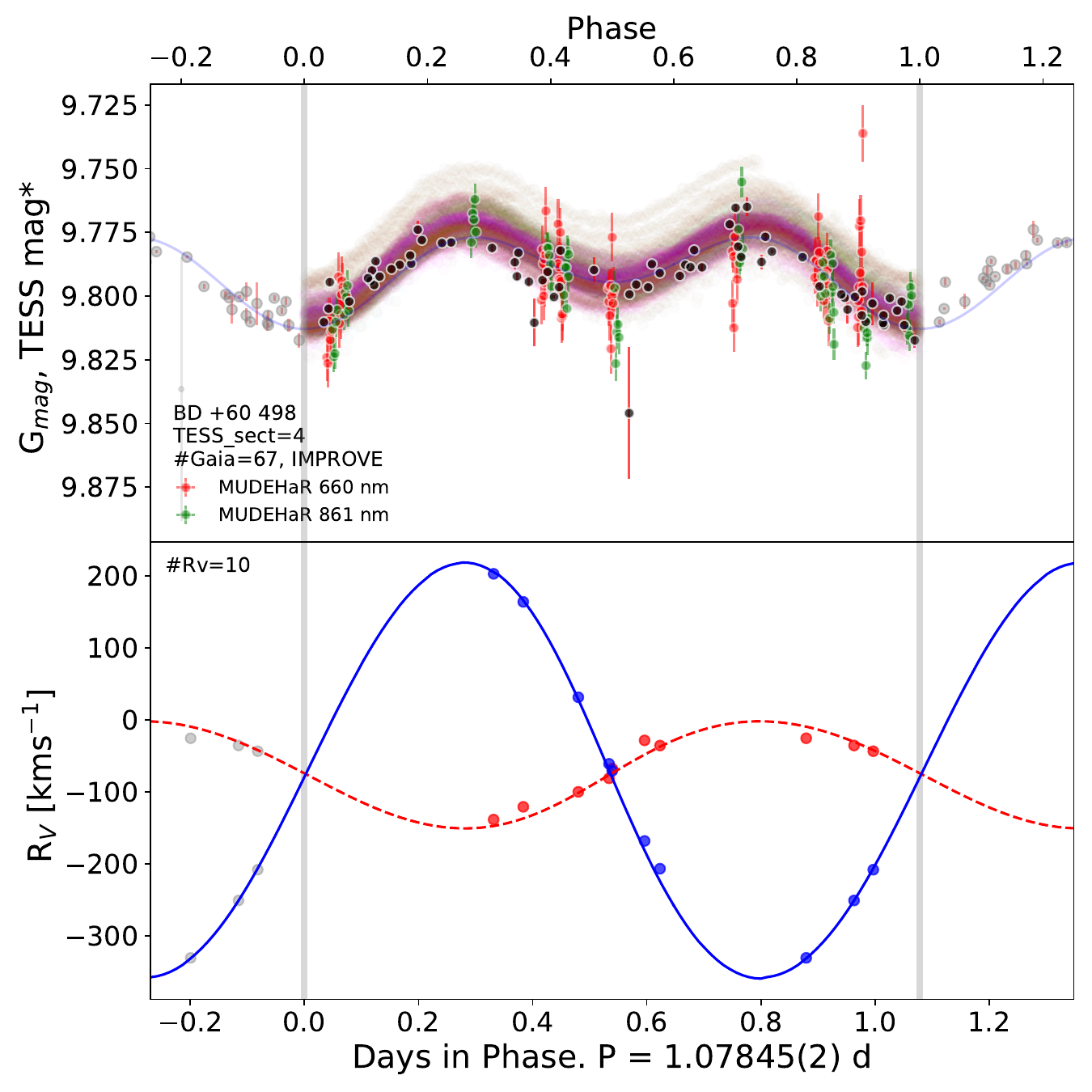}
    \caption{Same as Fig.~\ref{BD_+61_487_3p4537(5)_column}, for BD~$+$60~498, this time including MUDEHaR data.}
    \label{BD_+60_498_1p07845(2)_column.pdf}
\end{figure}

This system has been classified as being of O type by many authors since the 1950s, but to our knowledge, nobody had detected the presence of a weak B-type spectroscopic companion that makes the system an SB2E.
The secondary is barely detected at points close to quadrature, with a light ratio between the components of $\sim 10:1$ and a mass ratio of $\sim 3:1$. 
The light curve is sinusoidal-like with a standard deviation of 8-11~mmag depending on the filter but it does not appear in \citet{MaizApellaniz2023} because it has a six-parameter \textit{Gaia}~DR3 astrometric solution. 
Furthermore, the absorption lines in the LiLiMaRlin data do not show additional broadening, so the data point towards this system being a high-mass-ratio, low-inclination system without eclipses. 
Our calculated period is \num{1.07845}(2)~d, and SIMBAD lists a period with half the value (\num{0.539215}~d) from \textit{Gaia} epoch photometry, although not from \cite{Mowlavi2022}, but from \cite{GaiaCollaboration2022b}, likely coming from \citet{Rimoldini2022}, who classify this star as a BCEP ($\beta$~Cep). 
We have preliminary data from MUDEHaR on this star. Recalculating the period with MUDEHaR data showed negligible differences, indicating good compatibility.
Finally, \textit{Hipparcos} data were available; however, the photometric noise level was too high to allow for a reliable use.

As previously noted, the secondary component is scarcely discernible in the spectra and is evident only at complete quadrature (Fig.~\ref{BD+60498_1}). All evidence supports the hypothesis of a very low-mass companion or potentially a stripped star, due to the considerable mass disparity. However, this hypothesis is contested by the non-detection of the secondary component in the \ioni{He}{ii} lines, where such presence would be anticipated if a hot core were in the system. 
The \ioni{He}{ii}~$\lambda$5411 lines were employed to obtain the RVs of the primary component and \ioni{He}{i}~$\lambda$5875 line, for the secondary, which were subsequently processed with \textsc{unwind} to disentangle the spectra and obtain the final spectroscopic orbit and integrated with photometric data to construct and calibrate the \textsc{phoebe} model. 
We derive a spectral classification of O9.7~V~+~B4:~Vn, with the secondary having a significant uncertainty due to the poor S/N of the disentangled spectrum, as it contributes just 10\% of the total light of the system in the $B$ band (Fig.~\ref{spectra1}). 

In Fig.~\ref{BD_+60_498_1p07845(2)_column.pdf}, the alignment of the photometric observations and measured radial velocities with the adjusted \textsc{phoebe} model is presented. The dataset encompasses a substantial number of epoch photometry points from \textit{Gaia}, along with observations from four \textit{TESS} sectors (18, 58, 85, and 86). The parameters derived from the \textsc{phoebe} fit are enumerated in Table~\ref{PHOEBE_params}. These parameters indicate the complex scenario, characterized by a mass ratio of 3$\sim$4 while exhibiting similar radii for both components.
When compared with values reported in the literature, the effective temperatures derived in this study appear a bit lower those reported by \cite{Morrison1975} and \cite{Morossi1980}. Nonetheless, their systematically slightly higher values are likely attributable to the omission of blanketing effects in their atmospheric models \citep[see][]{Martins2005}.


\subsubsection{GLS~8673}


\begin{figure}
    \centering
    \includegraphics[width=0.49\textwidth]{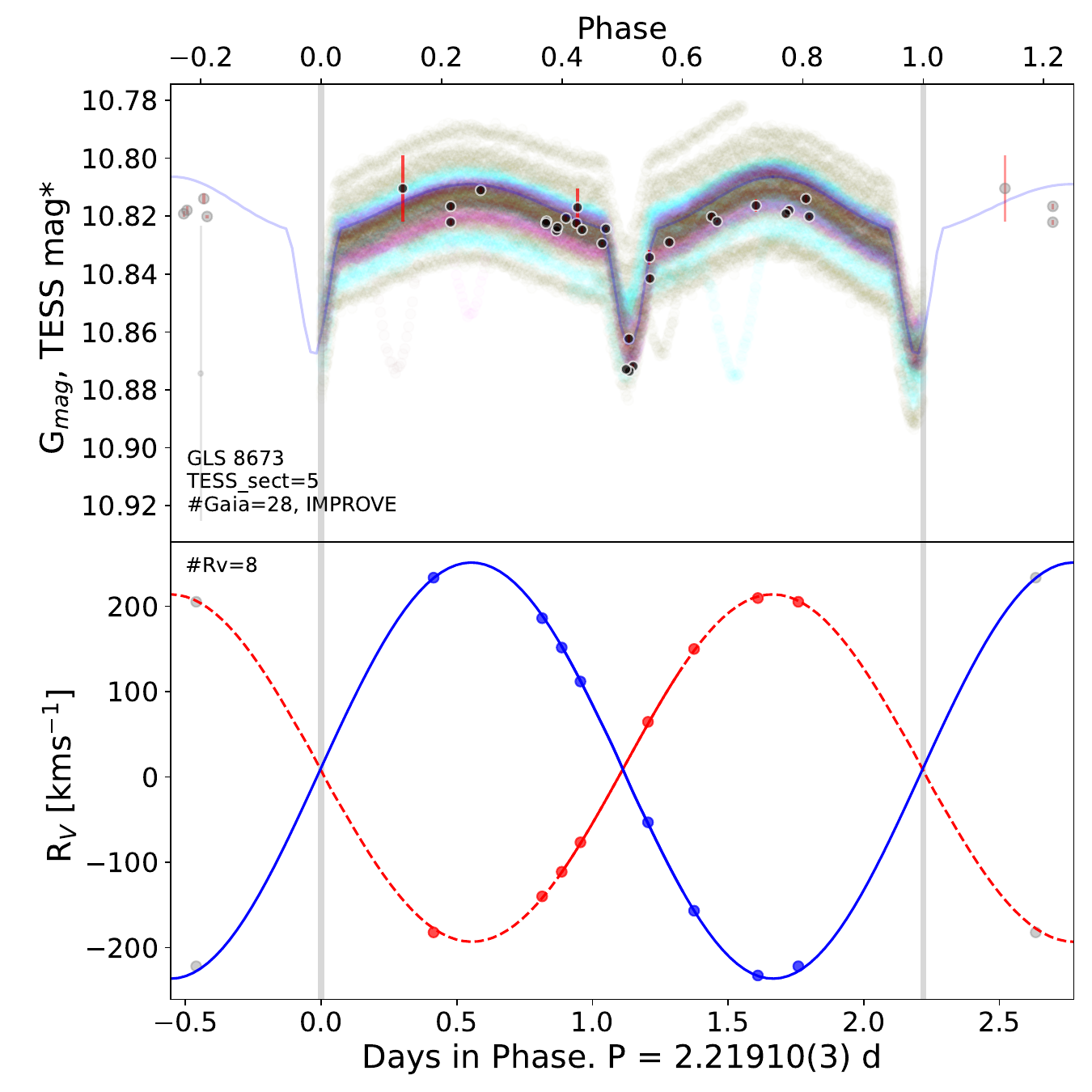}
    \caption{Same as Fig.~\ref{BD_+61_487_3p4537(5)_column}, for GLS~\num{8673}.}
    \label{ALS_8673_2p21910(3)_column.pdf}
\end{figure}


\begin{figure*}
    \centering
    \includegraphics[width=\textwidth]{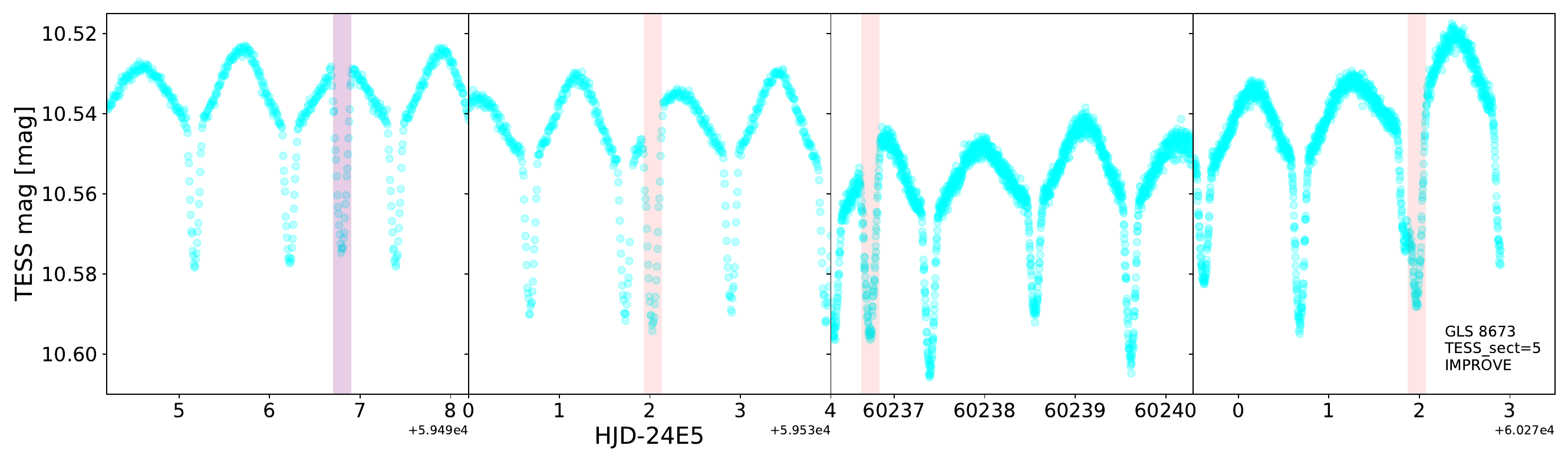}
    \caption{GLS~8673 \textit{TESS} data. The four eclipses found for the second system in the data, with period P=\num{35.235}~d. The first eclipse is used as HJD\_0 (band in purple), and then used as ephemerids to mark the expected eclipses (bands in orange).}
    \label{ALS_8673_LP}
\end{figure*}

GLS~8673\footnote{In the submitted ALS~III paper by Pantaleoni González et al. the ALS objects that have been confirmed to be Galactic massive stars have had their names changed from ALS (Alma Luminous Star) to GLS (Galactic Luminous Star) e.g. this object was previously known as ALS~8673.} is a poorly studied system that had been identified as an eclipsing binary before but for which we have found no mention in the literature about its spectroscopic binarity.
Our calculated photometric period is \num{2.21910}(3)~d. We consider it an improvement since in his article \cite{Mowlavi2022} provides a value of the period that is half or the one calculated in our work.
In the constructed light curve (Fig.~\ref{ALS_8673_2p21910(3)_column.pdf}), the second minimum/eclipse is poorly sampled in \textit{Gaia} (practically nonexistent), which could be be the source of the discrepancy.
Moreover, the light curve displays additional extra minima, which we identified as indicative of another binary system with a period of 35.235~d, see Fig.~\ref{ALS_8673_LP}. This factor has been considered in our assessment of the system.
Unfortunately, none of the data from \textit{Gaia} align with the anticipated eclipses corresponding to the secondary period. Additionally, although data from \textit{Hipparcos} were accessible, the level of photometric noise was excessively high, rendering them unsuitable for reliable analysis.

Upon visual analysis of the spectra (Fig.~\ref{ALS8673_5876_2Phases.pdf},~\ref{ALS8673_1}), the system's oscillatory period is distinctly evident in both the main \ioni{He}{i} lines and specifically the \ioni{He}{ii}~$\lambda$4686 line.
The similarity of the two components may result in ambiguity between the actual period and its half-value. Nonetheless, a minor discrepancy between these components corroborates the accuracy of our determined period.
In certain high-resolution spectra, where the separation of the two components' lines is pronounced, a faint component emerges between them. This component may potentially originate from the additional long period system. Nevertheless, it cannot be definitively claimed that an SB3 system is being observed. 
The \ioni{O}{iii}~$\lambda$5592 line remains undetectable, even at conjunction, likely due to dilution effects caused by moderately elevated rotation rates.
The \ioni{He}{i}~$\lambda$4922 and 5876 lines were utilized to derive radial velocity measurements, which were subsequently integrated with photometric data, facilitating the construction and calibration of the \textsc{phoebe} model.

Fitting the \textsc{phoebe} model in this case presented several complications.
Firstly, as previously noted, the \textit{TESS} data reveal four eclipses consistent with a second eclipsing binary system with a period of P=35.235 days (see Fig.\ref{ALS_8673_LP}). However, the morphology and timing of these eclipses—specifically, the similar phases and differing depths—make it unlikely that they originate from the same system as GLS~8673. In particular, if the 35-day signal were due to a secondary pair within the same multiple system, one would expect eclipses at complementary phases and possibly more symmetry in depth, unless substantial third light or other complexities were present.
Given that no corresponding eclipses are observed in \textit{Gaia} data, and the \textit{Hipparcos} photometry is too noisy for a reliable analysis, we consider the possibility that these eclipses emerge from a separate eclipsing binary along the same line of sight. Although this system is visible in \textit{TESS}, no spectroscopic signatures corresponding to a third component have been unambiguously identified, nor does the periodogram analysis (e.g., with \texttt{Pyriod}) recover the 35-day period in the radial velocity data.
If we hypothetically associate this long-period signal with a physically bound system composed of B1.5$\pm$0.5-type stars, the third-light contribution would be approximately 20\%. However, due to the lack of strong spectroscopic confirmation and the uncertainty in its nature, this contribution was not included in the \textsc{phoebe} model, and the fitted parameters for the short-period system may therefore be subject to small systematic biases. 

Focusing solely on the orbital behavior of the short-period system, no prior orbital solution for this star was found in the literature. The dataset includes a substantial number of epoch photometry points from \textit{Gaia} and observations from five \textit{TESS} sectors (43, 44, 45, 71, and 72).  
The spectroscopic orbital fit appears robust, and the orbit is considered credible. The parameters derived from the \textsc{phoebe} model are listed in Table~\ref{PHOEBE_params}, including an almost negligible eccentricity value of \(e = 0.006 \pm 0.006\). The masses obtained align well with theoretical expectations for the given spectral types, although the radii results are slightly lower than expected, from the expected 7 to approximately 5~\(R_{\odot}\) obtained. 
We emphasize that no third-light contribution was included in the fit and we caution that the masses and radii of could be affected at the 10–20\% level by unaccounted third light.
To the best of our knowledge, there have been no prior determinations of any of these parameters in the literature for this system to enable direct comparison.

The output spectra from \textsc{unwind} shows that GLS~8673 is a near-twin system, with both components contributing close to 50\% of the light and being classified as O9.7~V. The mass ratio is also the closest one to 1.0 in our sample.



\begin{table*}[!t]
\caption{Summary of star information, \textsc{phoebe} model fits, orbital fits, and spectral classifications, with the horizontal lines separating the groups.}
\label{PHOEBE_params}
\centering
\hspace*{-0.55cm}
\addtolength{\tabcolsep}{-4pt}
\begin{tabular}{lcccccccc}
\hline \hline
\noalign{\smallskip}
Star                 & BD~$+$61~487           & HDE~\num{255312}          & HD~\num{169727}        & BD~$-$13~4929          & BD~$+$60~498           & GLS~\num{8673}         & GLS~\num{19307}        & BD~$+$55~2840          \\
ALS ID               & GLS~\num{7485}         & GLS~\num{8891}            & GLS~\num{5046}         & GLS~\num{4913}         & GLS~\num{7269}         & GLS~\num{8673}         & GLS~\num{19307}        & GLS~\num{12685}        \\
\GGcor\ [mag]        & \num{9.6125(50)}       & \num{9.2330(60)}          & \num{8.9611(52)}       & \num{9.5989(50)}       & \num{9.7975(48)}       & \num{10.8277(48)}      & \num{11.4929(52)}      & \num{9.8474(50)}       \\
\BPRP\               & \num{1.0400}           & \num{1.1578}              & \num{1.1667}           & \num{1.0000}           & \num{0.7740}           & \num{0.5339}           & \num{1.8532}           & \num{0.6972}           \\
$\;\;$ [mag]         &                        &                           &                        &                        &                        &                        &                        &                        \\
\picor\ [mas]        & 0.180(22)              & 0.614(31)                 & 0.511(36)              & 0.619(89)              & 0.479(36)              & 0.258(27)              & 0.230(31)              & 0.296(29)              \\
$d$ [kpc]            & $5.88^{+0.90}_{-0.69}$ & $1.636^{+0.087}_{-0.079}$ & $1.99^{+0.15}_{-0.13}$ & $1.75^{+0.34}_{-0.24}$ & $2.13^{+0.18}_{-0.15}$ & $4.03^{+0.50}_{-0.40}$ & $4.54^{+0.77}_{-0.57}$ & $3.44^{+0.40}_{-0.32}$ \\
\midrule
\textsc{phoebe} type & Semi-Det.              & Overcont.                 & Overcont.              & Detached               & Semi-Det.              & Detached               & Detached               & Semi-Det.              \\
$P$ [d]              & 3.4537(5)              & 1.46972(5)                & 2.9302(2)              & 1.475230(1)            & 1.07845(2)             & 2.21910(3)             & 3.1240(5)              & 1.9631(2)              \\
Third light          & 0 (fix)                & 0 (fix)                   & 0 (fix)                & 64\%                   & 0 (fix)                & 0 (fix)                & 0 (fix)                & 0 (fix)                \\
$i$ [º]              & 30.5(30)               & 51.5(25)                  & 33.6(53)               & 81.7(20)               & 33.1(41)               & 68.5(15)               & 62.5(12)               & 42.0(32)               \\
sma [R$_{\odot}$]    & 36.1(32)               & 20.4(7)                   & 33(4)                  & 13.7(1)                & 14.7(14)               & 21.1(2)                & 33.0(3)                & 22.3(15)               \\
\Teff$_1$ [kK]       & 36.0(5)                & 39.0(50)                  & 35.2(12)               & 30.0(16)               & 31.0(22)               & 30.0(5)                & 34.1(9)                & 34.2(11)               \\
\Teff$_2$ [kK]       & 36.0(11)               & 31.0(10)                  & 34.0(21)               & 29.0(20)               & 18.0(28)               & 30.0(8)                & 32.0(13)               & 32.0(15)               \\
$M_1$ [M$_{\odot}$]  & 31.5(84)               & 29.5(31)                  & 28(12)                 & 8.869(68)              & 29.2(96)               & 13.91(43)              & 27.88(91)              & 22.8(42)               \\
$M_2$ [M$_{\odot}$]  & 19.3(52)               & 19.5(20)                  & 17(7)                  & 6.838(52)              & 7.0(23)                & 11.61(36)              & 17.3(56)               & 14.8(28)               \\
$R_1$ [R$_{\odot}$]  & 13.1(12)               & 9.2(3)                    & 14(2)                  & 4.66(2)                & 6.3(7)                 & 5.0(1)                 & 11.3(1)                & 9.3(6)                 \\
$R_2$ [R$_{\odot}$]  & 10.4(9)                & 7.7(3)                    & 12(2)                  & 3.78(2)                & 4.0(4)                 & 4.50(5)                & 10.3(2)                & 7.2(4)                 \\
$f_1$                & 1.09(7)                & 1.35(15)                  & 1.77(20)               & 0.065(69)              & $-$1.0(8)              & $-$2.24(6)             & $-$0.17(8)             & 1.07(7)                \\
$f_2$                & 0.86(7)                & ---                       & ---                    & $-$0.33(9)             & 1.06(75)               & $-$2.24(6)             & 0.7(1)                 & 0.82(10)               \\
\midrule
$T_0$ [d]            & 1.88(10)               & 1.1702(2)                 & 3.278(60)              & 1.4466(4)              & 1.4637(43)             & 1.65(15)               & 4.117 (fix)            & 2.26(32)               \\
$e$                  & 0.167(57)              & 0 (fix)                   & 0.149(17)              & 0 (fix)                & 0 (fix)                & \num{0.0067(59)}       & 0 (fix)                & \num{0.0098(56)}       \\
$\omega$ [º]         & 200(14)                & 90 (fix)                  & 126.7(72)              & 90 (fix)               & 90 (fix)               & 05(24)                 & 90 (fix)               & 212(59)                \\
$\gamma_1$ [km/s]    & $-$49.4(51)            & 0.12(27)                  & 17.8(10)               & 20.06(45)              & $-$74.2(29)            & 8.0(10)                & $-6$                   & $-$39.4(9)             \\
$\gamma_2$ [km/s]    & $-$55.3(53)            & $-$0.10(21)               & $-$0.9(13)             & 20.07(43)              & $-$60.8(76)            & 10.1(13)               & 8                      & $-$39.2(15)            \\
$K_1$ [km/s]         & 102.3(80)              & 213.60(35)                & 112.5(20)              & 201.79(84)             & 72.8(65)               & 203.5(12)              & 176(2)                 & 150.3(36)              \\
$K_2$ [km/s]         & 166.2(87)              & 322.97(33)                & 183.9(34)              & 261.72(69)             & 302.3(154)             & 243.7(16)              & 284(4)                 & 231.2(54)              \\
$q$                  & 0.615(54)              & \num{0.6613(13)}          & \num{0.6118(42)}       & 0.7710(38)             & 0.241(24)              & \num{0.8351(28)}       & 0.619(12)              & \num{0.6500(40)}       \\
\midrule
Sp. class 1          & Oe                     & O7 V((f))(n)z             & O7.5 Iabfp             & B0.2 V                 & O9.7 V                 & O9.7 V                 & O7 II(f)               & O7.5 V                 \\
Sp. class 2          & O                      & O9.5 V(n)                 & O7 Iabf                & B0.5 V                 & B4: V(n)               & O9.7 V                 & O8.5 II((f))           & ON8.5 IV               \\
$B$ flux fr. 1       & 50\%                   & 55\%                      & 25\%                   & 18\%                   & 90\%                   & 50\%                   & 70\%                   & 50\%                   \\
$B$ flux fr. 2       & 50\%                   & 45\%                      & 75\%                   & 18\%                   & 10\%                   & 50\%                   & 30\%                   & 50\%                   \\
\midrule
\multicolumn{9}{l}{$T_0$ is the Heliocentric Julian Date -\num{2456900}. No disentangling was performed for GLS~\num{19307}. The binary components for } \\
\multicolumn{9}{l}{BD~$-$13~4929 are Ba and Bb. The primary (A) has the 64\% of the remaining $B$ flux and has a spectral classification of O8 V.} \\
\multicolumn{9}{l}{The $\gamma$ uncertainties do not include the possible systematic effects of the cross-correlation with the \fastwind\ model.}
\end{tabular}
\addtolength{\tabcolsep}{4pt}
\end{table*}

\subsubsection{GLS~\num[detect-all]{19307}}


This system is the least studied in our sample, having been previously recognized as an eclipsing binary \citep{IJspeert2021} but with no previous spectroscopic study. We have not been able to find even a spectroscopic classification in the litearture.

\begin{figure}
    \centering
    \includegraphics[width=0.5\textwidth]{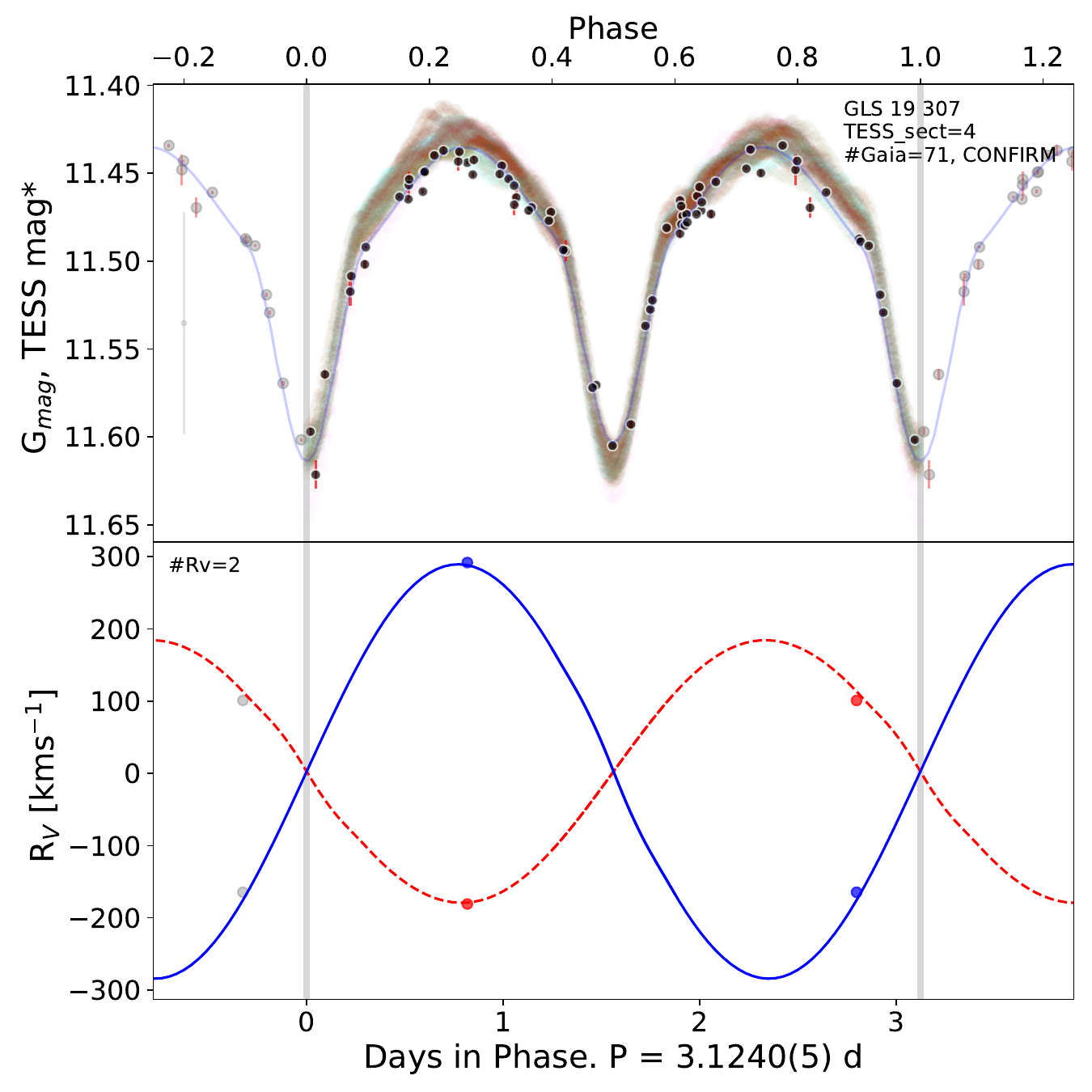}
    \caption{Same as Fig.~\ref{BD_+61_487_3p4537(5)_column}, for GLS~\num{19307}.}
    \label{ALS_19_307_3p1240(5)_column.pdf}
\end{figure}

Our calculated photometric period is \num{3.1240}(5)~d, consistent with the value reported by \citet{Mowlavi2022} and close to the previous value of \num{3.1237}~d indicated in \cite{Ren2021}.  
The light curve shows an excellent fit (Fig.~\ref{ALS_19_307_3p1240(5)_column.pdf}), clearly indicating the presence of eclipses.  
With only two spectra available, the radial velocities available do not properly complement the photometric orbit with spectroscopic data and we cannot apply \textsc{unwind}. However, we were able to confirm its SB2 classification (Fig.~\ref{GLS19307_5876.pdf}), derive a spectral classification from a LiLiMaRlin spectrum as O7~II(f)~+~O8.5~II((f)) using MGB (Fig.~\ref{spectra3}), and present a preliminary orbital solution, the first documented in the literature to our knowledge. 
In this case, the \ioni{O}{iii}~$\lambda$5592 line is clearly visible for the luminous companion, and barely detectable for the secondary companion.
Radial velocity measurements were obtained using the \ioni{He}{i}~$\lambda$5875 lines, which were then combined with photometric data to develop and calibrate the \textsc{phoebe} model.

In our \textsc{phoebe} fit (Fig.~\ref{ALS_19_307_3p1240(5)_column.pdf}), the observations and calculated models are in display for comparison. The parameters from the fit are displayed in Table~\ref{PHOEBE_params}. 
Photometric data are sourced from both \textit{Gaia} and \textit{TESS} sectors 14, 41, 54, and 81. 
With RV information available from only two spectra, both data points fit perfectly with the \textsc{phoebe} model; however, the resulting orbital solution remains highly preliminary.
This system remains poorly studied (less than 10 articles since the year 1900) and presents an interesting target for future observations aimed at determining a better orbital solution and conducting a quantitative spectroscopic analysis using atmospheric model fitting.


\subsubsection{BD~$+$55~2840 = GLS~\num[detect-all]{12685}}

\begin{figure}
    \centering
    \includegraphics[width=0.5\textwidth]{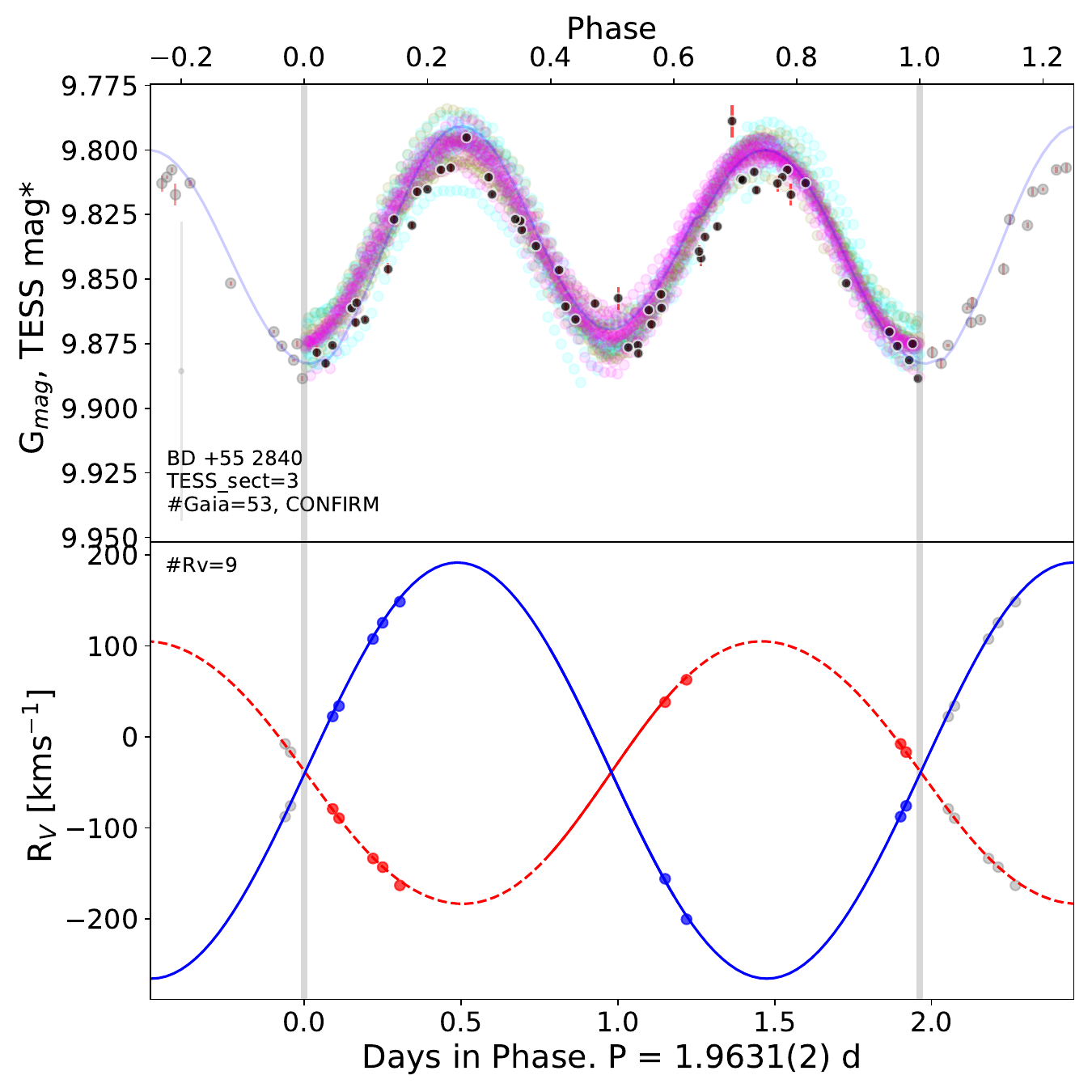}
    \caption{Same as Fig.~\ref{BD_+61_487_3p4537(5)_column}, for BD~$+$55~2840.}
    \label{BD_+55_2840_1p9631(2)_column.pdf}
\end{figure}

In \citet{Maizetal16} this system was classified as O7.5~V(n), with the (n) suffix indicating broadened lines, but the system was not resolved as an SB2. Nevertheless, it was previously known to be an eclipsing binary and our calculated period is \num{1.9631}(2)~d, similar to the \citet{Mowlavi2022} value. 
The light curve is sinusoidal-like with a standard deviation of 24-27~mmag depending on the filter but it does not appear in \citet{MaizApellaniz2023} because it has a six-parameter \textit{Gaia}~DR3 astrometric solution.
Nevertheless, the light curve presented in Fig.~\ref{BD_+55_2840_1p9631(2)_column.pdf} shows a robust photometric fit. 

When plotting the available spectra for this system (Fig.~\ref{BD+552850_1}), the components appear moderately separated. Both the He\,\textsc{i} and He\,\textsc{ii} lines show similar widths for each component. Additionally, the \ioni{O}{iii}~$\lambda$5592 line seems to be associated with the star that exhibits weaker spectral lines. The He\,\textsc{i} 4922\,\AA\ lines were used to derive the initial radial velocity measurements prior to the subsequent analysis with \textsc{unwind} and the posterior combination with photometric data to construct and fit the \textsc{phoebe} model.

The disentangled spectra yield a first-time classification as an SB2E of O7.5~V~+~ON8.5~IV. We point out that the light in the $B$ band is evenly divided between the two components but the primary is significantly more massive. The secondary is nitrogen-enriched (Fig.~\ref{spectra1}), so it is possible that mass has been transferred from the secondary to the primary, leaving processed material exposed in the first one of those. 

Figure~\ref{BD_+55_2840_1p9631(2)_column.pdf} shows the PHOEBE model fit to our measurements (model parameters are listed in Table~\ref{PHOEBE_params}). The \textit{TESS} sectors used were 16, 17, and 24. 
The fit to the radial velocity data is remarkably good and, to the best of our knowledge, this is the first presentation of the orbital solution for this system in the literature.
The PHOEBE model parameters lie near the expected boundaries for the given spectral type.

\subsection{Systems with previous determinations in the literature}

\subsubsection{HD~\num[detect-all]{1810}~A = V745~Cas~A = GLS~\num[detect-all]{6159}}


\begin{figure}
    \centering
    \includegraphics[width=0.49\textwidth]{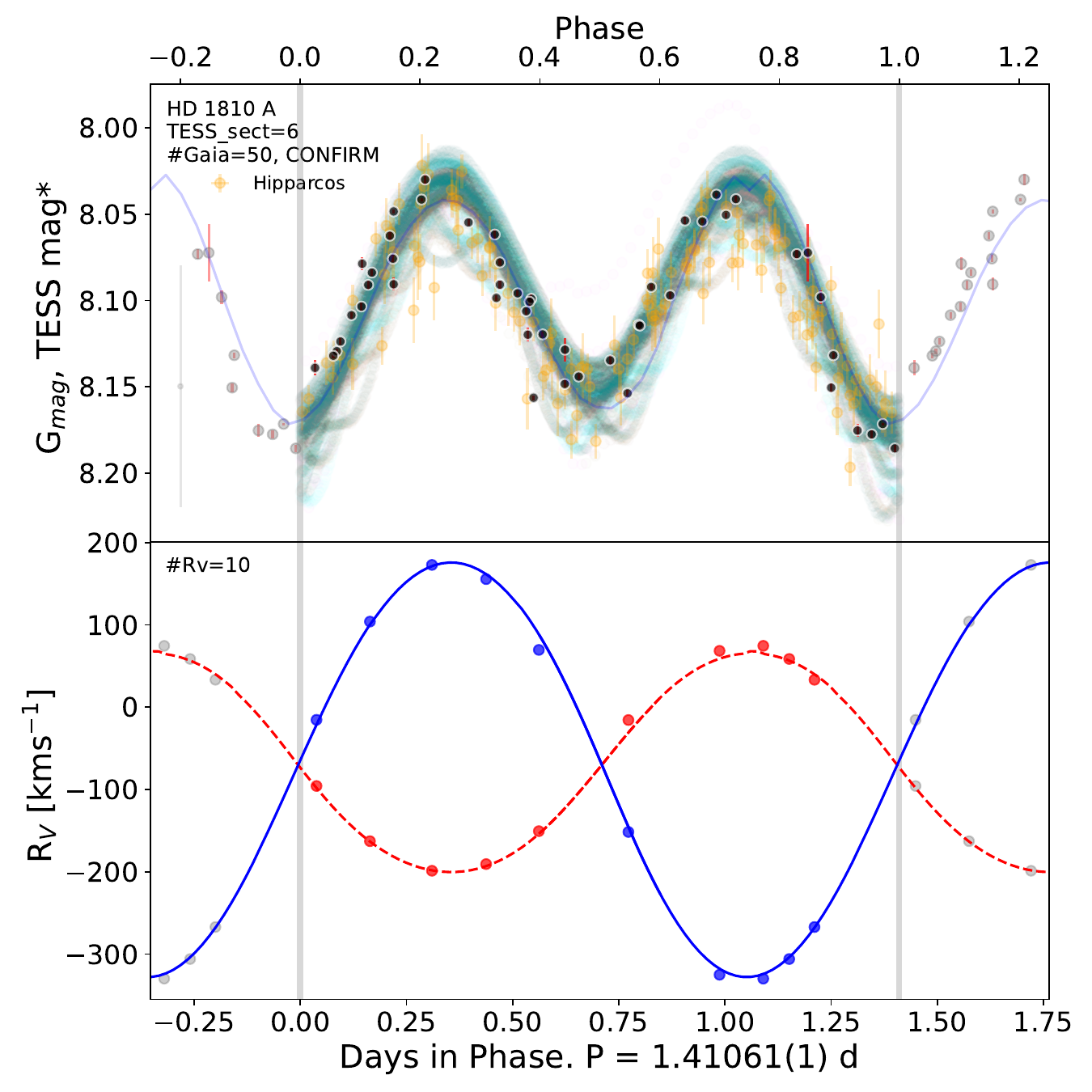}
    \caption{Same as Fig.~\ref{BD_+61_487_3p4537(5)_column}, for HD~\num{1810}~A, this time including \textit{Hipparcos} data.}
    \label{HD_1810_A_1p41061(1)_column.pdf}
\end{figure}

The literature period \citep{Alfonso-Garzon2012a,cakirli2014,Mowlavi2022} is similar to that calculated in this work, \num{1.41061}(1)~d.
This system has been analyzed primarily to evaluate the reliability of our methodology against previous studies in the literature. To this end, we have incorporated newly available data from \textit{Gaia} and \textit{TESS}, while also including \textit{Hipparcos} data, which have been extensively used in the literature to determine the period of the system.

The spectral analysis reveals the presence of two moving components for the \ioni{He}{i} lines, and \ioni{He}{ii} (and potentially \ioni{O}{iii}) lines for the more luminous component, suggesting the existence of at least one O-type star within the system (see Fig.~\ref{HD1810_5411}). The rotational velocity of the primary component is high, albeit not extreme, possibly indicating tidal interactions and synchronization effects. Radial velocity measurements were initially derived using the \ioni{He}{ii}~$\lambda$4686 line for the primary component (which is almost single, i.e. not contaminated by the other component) and \ioni{He}{i}~$\lambda$4922 line, for the secondary, and subsequently processed with \textsc{unwind} and integrated with photometric data to develop and calibrate the \textsc{phoebe} model.

In Fig.~\ref{HD_1810_A_1p41061(1)_column.pdf}, we present our orbital fit using the \textsc{phoebe} model, incorporating \textit{Hipparcos} data.
The new photometric data is obtained from both \textit{Gaia} and \textit{TESS} sectors 17, 18, 24, 58, 78, and 85.
The fit of the radial velocity curve, given the number of available spectra, is excellent.
HD~\num[detect-all]{1810}~A possesses a companion, designated HD~\num[detect-all]{1810}~B, situated at 9\farcs5 with a \textit{Gaia} magnitude difference of 2.2. We estimate that, in the context of \textit{TESS} measurements, this companion contributes approximately 12\% of the total flux. Owing to this nominal contribution - and the excellent fit of the light curve to the model - we conclude it is unnecessary to incorporate it as third light in the \textsc{phoebe} model.

\cite{cakirli2014} previously calculated a highly precise spectroscopic orbit, classifying the system as a contact binary\footnote{Note the typo in the title of that paper. The Latin genitive of \textit{Cassiopeia} is \textit{Cassiopeiae}, not \textit{Cassiopean}.}. Their analysis derived the orbital period using \textit{Hipparcos} data and determined the atmospheric parameters of the components by combining the spectra for two standard B-type stars that best matched the system’s composite spectrum.
Table~\ref{HD1810A_table} presents the parameters derived from our \textsc{phoebe} fit, in conjunction with the orbital results presented in \cite{cakirli2014}. A comparative analysis shows consistency across the majority of parameters. Nonetheless, we characterize the system as 'overcontact but not in thermal contact, and our calculated orbit exhibits a non-negligible eccentricity ($e$=0.0216$\pm$0.0027). Our disentangling-derived spectral classification of O9.5~V(n)~+~B1~V is more precise than that of \citet{cakirli2014} and the luminosity classification of the secondary of V is in better agreement with the measured flux and mass ratios, as those authors claimed it was a giant.

Ultimately, the orbital results reported in \cite{cakirli2014} were utilized in \cite{Vrancken2024} to explore potential temporal variations in the period. Although a relatively high $\dot{P}$ value was identified, it was accompanied by an uncertainty of nearly the same magnitude. In light of the slight difference in the period values between our findings and those documented in \cite{cakirli2014}, alongside the strong concordance of our period with all photometric data, we discern no evidence of period variation. Moreover, the integration of their tabulated $\dot{P}$ value into our \textsc{phoebe} fit does not enhance the agreement between model and observations.


\begin{table}[!t]
\caption{Information for HD~\num{1810}~A and comparison of results between this work and \citet{cakirli2014} (\citealt{Fabry2025} for R/RL).}
\label{HD1810A_table}
\centering
\addtolength{\tabcolsep}{-2pt}
\begin{tabular}{lcc}
\hline \hline
\noalign{\smallskip}
\GGcor\ [mag]        & \multicolumn{2}{c}{\num{8.1031(52)}}                 \\
\BPRP\ [mag]         & \multicolumn{2}{c}{\num{0.1024}}                     \\
\picor\ [mas]        & \multicolumn{2}{c}{0.74(39)}                         \\
$d$ [kpc]            & \multicolumn{2}{c}{$7.9^{+8.9}_{-4.6}$}              \\
\midrule
                     & This work             & \citet{cakirli2014}          \\
\midrule
\textsc{phoebe} type & Overcontact           & Contact                      \\
$P$ [d]              & \num{1.41061(1)}      & 1.41057(5)                   \\
Third light          & 0 (fix)               & 0 (fix)                      \\
i [º]                & 39.0(44)              & $\sim$43.7                   \\
sma [R$_{\odot}$]    & 17.9(17)              & 17.3(1)                      \\ 
\Teff$_1$ [kK]       & 32.0(13)              & 30.0(10)                     \\
\Teff$_2$ [kK]       & 26.0(13)              & 25.5(3)                      \\
$M_1$ [M$_{\odot}$]  & 23.1(66)              & 18.31(51)                    \\
$M_2$ [M$_{\odot}$]  & 12.4(35)              & 10.47(28)                    \\
$R_1$ [R$_{\odot}$]  & 8.0(8)                & 6.94(7)                      \\
$R_2$ [R$_{\odot}$]  & 6.2(6)                & 5.35(5)                      \\
$f_1$ or R/RL*       & 1.46(8)               & 1.00                         \\
$f_2$ or R/RL*       & -                     & 0.99                         \\
\midrule
$T_0$ [d]            & \num{2456900.418(21)} & \num{2455222.1234(2)}        \\
$\omega$ [º]         & 357.2(53)             & 90 (fix)                     \\
$e$                  & 0.0216(27)            & 0 (fix)                      \\
$\gamma_1$ [km/s]    & $-$66.32(42)          & $-$16(2)                     \\
$\gamma_2$ [km/s]    & $-$74.11(64)          & $-$16(2)                     \\
$K_1$ [km/s]         & 137.52(41)            & 156(2)                       \\
$K_2$ [km/s]         & 255.24(37)            & 273(3)                       \\
$q$                  & 0.5388(18)            & 0.5714(96)                   \\
\midrule
Sp. class 1          & O9.5~V(n)             & O9.5-B0.5~V                  \\
Sp. class 1          & B1 V                  & B1-2~III                     \\
$B$ flux fr. 1       & 75\%                  & ---                          \\
$B$ flux fr. 2       & 25\%                  & ---                          \\
\hline
\multicolumn{3}{l}{$T_0$ is the Heliocentric Julian Date.}
\end{tabular}
\addtolength{\tabcolsep}{2pt}
\end{table}

\subsubsection{DN~Cas = GLS~\num[detect-all]{7142}}


\begin{table}[!t]
\caption{Information for DN~Cas and comparison of results between this work and \cite{Bakics2016}}
\label{PHOEBE_DNCas}
\centering
\addtolength{\tabcolsep}{-2pt}
\begin{tabular}{lcc}
\hline \hline
\noalign{\smallskip}
\GGcor\ [mag]        & \multicolumn{2}{c}{\num{9.6799(54)}}                 \\
\BPRP\ [mag]         & \multicolumn{2}{c}{\num{1.0113}}                     \\
\picor\ [mas]        & \multicolumn{2}{c}{0.506(24)}                        \\
$d$ [kpc]            & \multicolumn{2}{c}{$1.993^{+0.099}_{-0.090}$}        \\
\midrule
                     &  This work            & \cite{Bakics2016}            \\
\midrule
\textsc{phoebe} type &  Detached             & Detached                     \\
$P$ [d]              & 2.31095(7)            & 2.310951(1)                  \\
Third light          & 0 (fix)               & 0                            \\
$i$ [º]              & 79.2(20)              & 77.2(2)                      \\
sma [R$_{\odot}$]    & 23.8(2)               & 23.53(5)                     \\
\Teff$_1$ [kK]       & 32.6(6)               & 32.1(10)                     \\
\Teff$_2$ [kK]       & 29.6(7)               & 28.2(11)                     \\
$M_1$ [M$_{\odot}$]  & 19.55(39)             & 19.04(7)                     \\
$M_2$ [M$_{\odot}$]  & 14.15(28)             & 13.73(5)                     \\
$R_1$ [R$_{\odot}$]  & 7.5(1)                & 7.22(6)                      \\
$R_2$ [R$_{\odot}$]  & 4.82(9)               & 5.79(6)                      \\
$f_1$                & 0.58(9)               & $-$0.10(7)                   \\
$f_2$                & $-$2.53(9)            & $-$0.22(9)                   \\
\midrule
$T_0$ [d]            & \num{2456901.9923(5)} & \num{2456287.2699}           \\
$\omega$ [º]         & 90 (fix)              & 90 (fix)                     \\
$e$                  & 0 (fix)               & 0 (fix)                      \\
$\gamma_1$ [km/s]    & $-$47.40(28)          & $-$45.5(3)                   \\
$\gamma_2$ [km/s]    & $-$47.68(28)          & $-$45.5(3)                   \\
$K_1$ [km/s]         & 214.60(40)            & 211(3)                       \\
$K_2$ [km/s]         & 296.35(34)            & 292(5)                       \\
$q$                  & 0.7242(16)            & 0.722(2)                     \\
\midrule
Sp. class 1          & O8.5~V(n)             & B0 V                         \\
Sp. class 1          & B0 V                  & B1 V                         \\
$B$ flux fr. 1       & 75\%                  & ---                          \\
$B$ flux fr. 2       & 25\%                  & ---                          \\
\hline
\multicolumn{3}{l}{$T_0$ is the Heliocentric Julian Date.}
\end{tabular}
\addtolength{\tabcolsep}{2pt}
\end{table}

\citet{MaizApellaniz2019} classified this SB2E+Ca system as O8.5 V +B0.2~V. There are likely two additional stars in the system, one detected by light-time effects by \citet{Bakics2016} and the other one with lucky imaging by \citet{MaizApellaniz2019} using AstraLux \citep{MaizApellaniz2010, MaizApellaniz2021_lucky}. Our calculated period is \num{2.31095}(7)~d, similar to the values found in the literature \citep{Kreiner2004,Bakics2016,Mowlavi2022}.
In this instance, the intention of our results is not to refine the orbits or values established in prior studies, as they are entirely equivalent and consistent. The aim of our study is to validate our methodology by comparing it with existing results in the literature. 
In future research, we intend to use this calculated orbit to examine the system's atmospheric parameters. This approach will enable us to compare our findings with those reported by \citet{Bakics2016}, based on fits to Kurucz model spectra.

The separation of the spectral lines of the two components is pronounced, maintaining full independence at quadrature (Fig.~\ref{DN_Cas_1}). This condition offers an optimal framework for atmospheric analyses we intend to pursue in a subsequent article, thereby justifying our objective to derive an independent orbit based on our data. 
The \ion{He}{II} $\lambda$~4686 line is barely visible for the companion  (Fig.~\ref{BD+60470_4686}), reinforcing the B0 classification of the secondary component.
The \ioni{O}{iii} line is exclusively linked to the O-type star, though it presents a weak stationary component, similar to the case of BD~-13~4929. Radial velocity measurements were initially derived from the \ioni{He}{i}~$\lambda$5875 lines, which were then processed with \textsc{unwind} and integrated with photometric data for the purpose of developing and calibrating the \textsc{phoebe} model.

\begin{figure}
    \centering
    \includegraphics[width=0.49\textwidth]{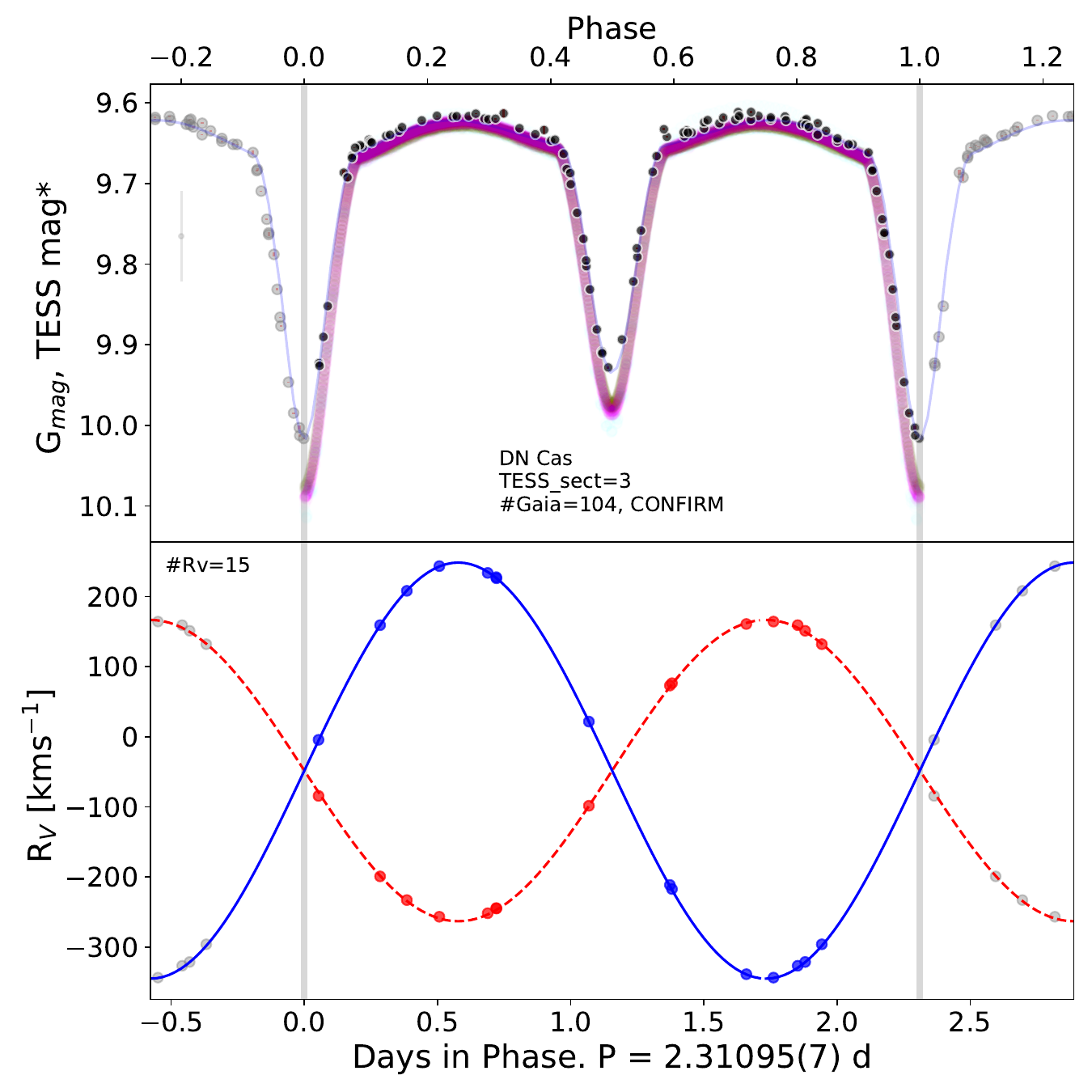}
    \caption{Same as Fig.~\ref{BD_+61_487_3p4537(5)_column}, for DN~Cas.}
    \label{DN_Cas_2p31095(7)_column.pdf}
\end{figure}

Figure~\ref{DN_Cas_2p31095(7)_column.pdf} presents our orbital solution, which demonstrates an excellent fit for both photometry and the radial velocities of both components. 
\textit{TESS} sectors used in this figure are 18, 85, and 86.
In a previous thorough and comparable study, \citet{Bakics2016} utilized their own photometric and spectroscopic observations and achieve an equally robust fit. They also suggested the presence of at least one physically bound, distant companion with a very low mass and an extended orbital period of \(P=42\) years.  
Table~\ref{PHOEBE_DNCas} lists the parameters derived from our \textsc{phoebe} analysis alongside those obtained by \citet{Bakics2016}. Most parameters are compatible but we point a significant difference: our spectral classification derived from disentangled spectra yields O8.5~V(n)~+~B0~V, significantly earlier than the spectral types from \citet{Bakics2016}, which were both of B type. The system is detached, with zero eccentricity and an inclination close to \(i\sim78^\circ\). Differences are minimal, though we note a slightly lower radius for the secondary star, which still aligns with expectations for its spectral type.  

\citet{Bakics2016} found consistency between these orbital parameters and those derived through quantitative spectral analysis using Kurucz models. In the future, we plan to perform a similar analysis using {\sc fastwind} \citep{Santolaya-Rey1997, Puls2005}, which will allow us to explore the potential advantages of employing an atmospheric code specifically designed for this type of star.

\section{Conclusions}\label{conclusions}

Within the framework of the MONOS project, we have presented new orbital solutions for ten spectroscopic O-type multiple systems combining \textit{Gaia} DR3 epoch photometry, \textit{TESS} light curves, and high-resolution spectroscopy. Eight of these systems lacked previous published orbits, while the remaining two have been refined with additional data. 
This number of new orbital determinations significantly enhance the sample of well-characterized massive binaries \citep[e.g.,][]{Cester1979,Stic97,Masoetal01,Khaliullin2010,Mayer2014,Malkov2020,MaizBarb20,IJspeert2021,Southworth2022,Ansietal23,Cakirli2023}, particularly in the short-period regime, where tidal interactions and mass transfer effects are dominant, and the overcontact binaries \citep{Sen2022,Abdul-Masih2025}. 

The combined use of multi-epoch photometry from \textit{Gaia} and \textit{TESS}, together with spectroscopic coverage, enables scalable and effective orbital characterization of massive binaries. 
Notable discoveries include the first confirmed Oe+O spectroscopic binary (BD~+61~487) and a system composed of overcontact O-type supergiants in an eccentric orbit (HD~169~727), both indicative of advanced binary interaction and evolutionary processes.
Notably, one binary displays a period exactly half the value reported by \citet{Mowlavi2022}, while another system reveals a second eclipsing binary with a 35-day period, hidden behind a more dominant 2-day eclipse. These cases demonstrate the inherent limitations of semi-automatic classification pipelines \citep[e.g.,][]{Rimoldini2022} and underscore the continued need for manual inspection and critical evaluation of light curves. 
All the derived solutions provide evidence for the significance and complexity of hierarchical multiples among O-type stars \citep[e.g.,][]{Sanaetal14,SimonDiaz2015b,Martetal17b}.

Each short-period SB2 or eclipsing system studied here represents a substantial addition to the still-scarce population of well-characterized massive binaries. These systems are especially valuable for probing mass transfer, tidal interaction, and evolutionary effects, as they provide direct dynamical constraints on masses, inclinations, and orbital configurations. Our sample includes the first orbital characterizations of several systems below the 3-day period regime, which is particularly sensitive to such interactions.

In addition to the new orbital fits, this work serves as a methodological benchmark, demonstrating the power of combining \textit{Gaia} epoch photometry, \textit{TESS} light curves, and preliminary MUDEHaR photometry to derive and validate photometric periods. All orbital and stellar parameters were obtained using the \texttt{phoebe} modeling framework and are broadly consistent with spectral classifications and theoretical expectations, including short periods ($P<3$~days), high mass ratios, and semi-detached or overcontact configurations. 
As more spectroscopic and photometric data become available, particularly for long-period systems, this integrated strategy can be scaled to achieve full phase coverage and improved orbital solutions across a wider parameter space.

This work provides a robust framework for future large-scale analyses of massive binaries, and lays the foundation for subsequent quantitative spectroscopic studies.
These will allow us to derive accurate stellar parameters for SB2 systems, enabling comparisons with the extensive datasets for single and SB1 stars \citep[e.g.,][]{SimonDiaz2011,Holgado2020,Holgado2022, Britetal23}.
Such comparisons are critical for advancing our understanding of how binarity affects stellar structure and evolution.
Finally, the techniques and results developed in this study are directly applicable to southern hemisphere O-type systems through the upcoming MOSOS project.
Its extension will support the MONOS project’s aim of building a complete and statistically robust census of massive multiples in the Galaxy.


\begin{acknowledgements}

We are extremely grateful to the referee of this paper for his valuable comments and for providing a constructive and motivating report to improve this work. 

G.H. acknowledges support from the State Research Agency (AEI) of the Spanish Ministry of Science and Innovation (MICIN) and the European Regional Development Fund, FEDER under grants LOS MÚLTIPLES CANALES DE EVOLUCIÓN TEMPRANA DE LAS ESTRELLAS MASIVAS/ LA ESTRUCTURA DE LA VIA LACTEA DESVELADA POR SUS ESTRELLAS MASIVAS with reference PID2021-122397NB-C21 / PID2022-136640NB-C22 / 10.13039/501100011033. 
G.~H. and J.~M.~A. acknowledge support from the Spanish Government Ministerio de Ciencia e Innovaci\'on and Agencia Estatal de Investigaci\'on (\num{10.13039}/\num{501100011033}) through grants PGC2018-0\num{95049}-B-C22 and PID2022-\num{136640}NB-C22 and from the Consejo Superior de Investigaciones Cient\'ificas (CSIC) through grant 2022-AEP~005. 
R.~G. acknowledges support from grant PID-2024 UNLP G189.


The project leading to this application has received funding from European Commission (EC) under Project OCEANS - Overcoming challenges in the evolution and nature of massive stars, HORIZON-MSCA-2023-SE-01, No G.A 101183150.
Funded by the European Union. Views and opinions expressed are however those of the author(s) only and do not necessarily reflect those of the European Union or (name of the granting authority). Neither the European Union nor the granting authority can be held responsible for them.

This work has made use of data from the European Space Agency (ESA) mission {\it Gaia} (\url{https://www.cosmos.esa.int/gaia}), processed by the {\it Gaia} Data Processing and Analysis Consortium (DPAC, \url{https://www.cosmos.esa.int/web/gaia/dpac/consortium}). Funding for the DPAC has been provided by national institutions, in particular the institutions participating in the {\it Gaia} Multilateral Agreement.

It used observations made with the JAST/T80 telescope/s at the Observatorio Astrofísico de Javalambre, in Teruel, owned, managed and operated by the Centro de Estudios de Física del Cosmos de Aragón. 
The initial part of the reduction and astrometric calibration was done by the OAJ Data Processing and Archiving Unit (UPAD) and the main part of the reduction and calibration was done at the Centro de Astrobiología.

Based on observations made with the Nordic Optical Telescope, operated by NOTSA, and the Mercator Telescope, operated by the Flemish Community, both at the Observatorio del Roque de los Muchachos (La Palma, Spain) of the Instituto de Astrofísica de Canarias.  

The work reported on in this publication has been partially supported by COST Action CA18104: MW-Gaia.


This research has made use of the SIMBAD database, operated at CDS, Strasbourg, France. 


\end{acknowledgements}


\bibliographystyle{aa}
\bibliography{sample}


\begin{appendix}
\section{Figures}\label{FiguresIndiv}

\subsection{Spectra}\label{Spectra}

In this section, we present three types of spectral figures. The first set shows the complete series of spectra available for selected stars of particular interest, as discussed in the main text, along with their corresponding orbital phases. Various diagnostic lines have been employed, as detailed in Sect.~\ref{Results}.

The second set includes the spectra used for spectral classification, displaying all available spectral and diagnostic lines to emphasize relevant features used for classification

Finally, the third set presents a series of figures illustrating the quality of the spectral disentangling achieved with the {\sc UNWIND} program for the different components.

\begin{figure}
    \centering
    \includegraphics[width=0.5\textwidth]{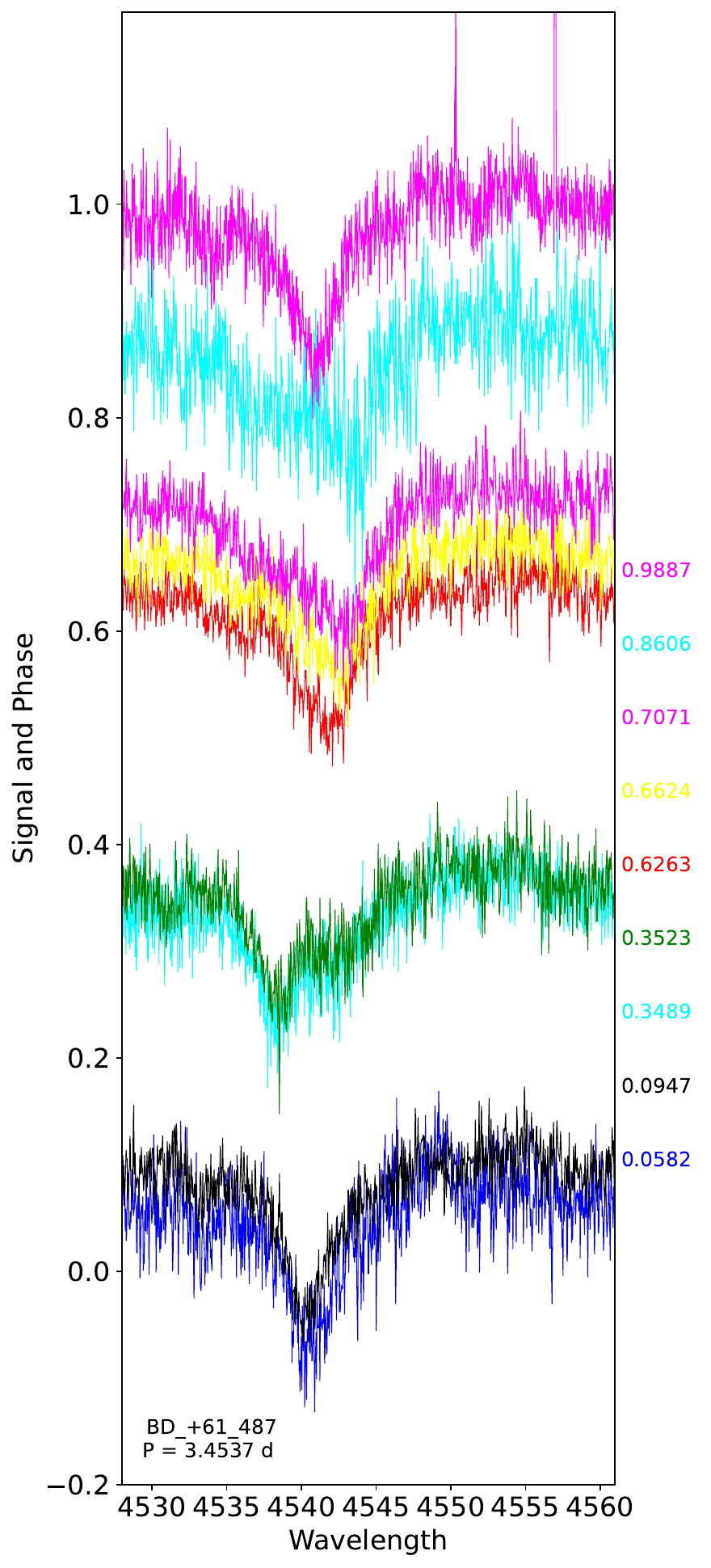}
    \caption{The composite spectrum of BD~$+$61~487 adjusted in height to the phase of each of the observations, for the \ion{He}{I} $\lambda$~4541 line. The color is assigned randomly, and the orbital phase of each spectrum—matching the same color—is displayed on the right side. This figure in combination with Fig.~\ref{BD+61487_4471_in.pdf} show the Oe+O nature of the system.}
    \label{BD+61487_4541}
\end{figure}

\begin{figure}
    \centering
    \includegraphics[width=0.5\textwidth]{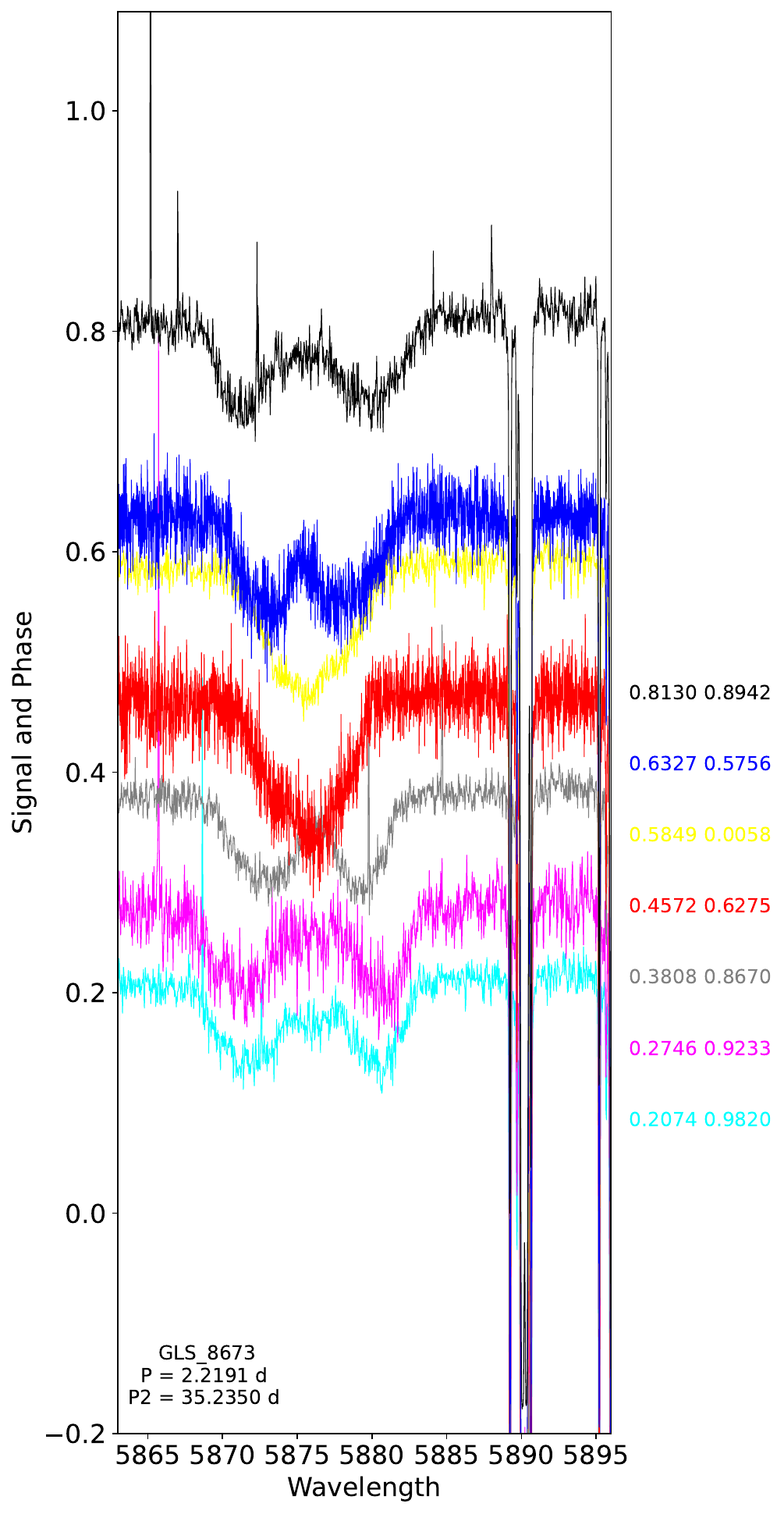}
    \caption{Same as Fig.~\ref{BD+61487_4541} for GLS~8673 and for the \ion{He}{I} $\lambda$~5875 line. 
    For this system, the orbital phase of each spectrum—matching the same color—is displayed on the right side with respect to the two periods derived from photometry.
    We note a marginal feature between the spectral lines of the short-period binary system, in two spectra (cyan and pink) which were taken during quadratures of the short-P eclipsing and coincident with the conjunction of the 35.2-d one. But, they should be confirmed with more data.}
    \label{ALS8673_5876_2Phases.pdf}
\end{figure}

\begin{figure}
    \centering
    \includegraphics[width=0.5\textwidth]{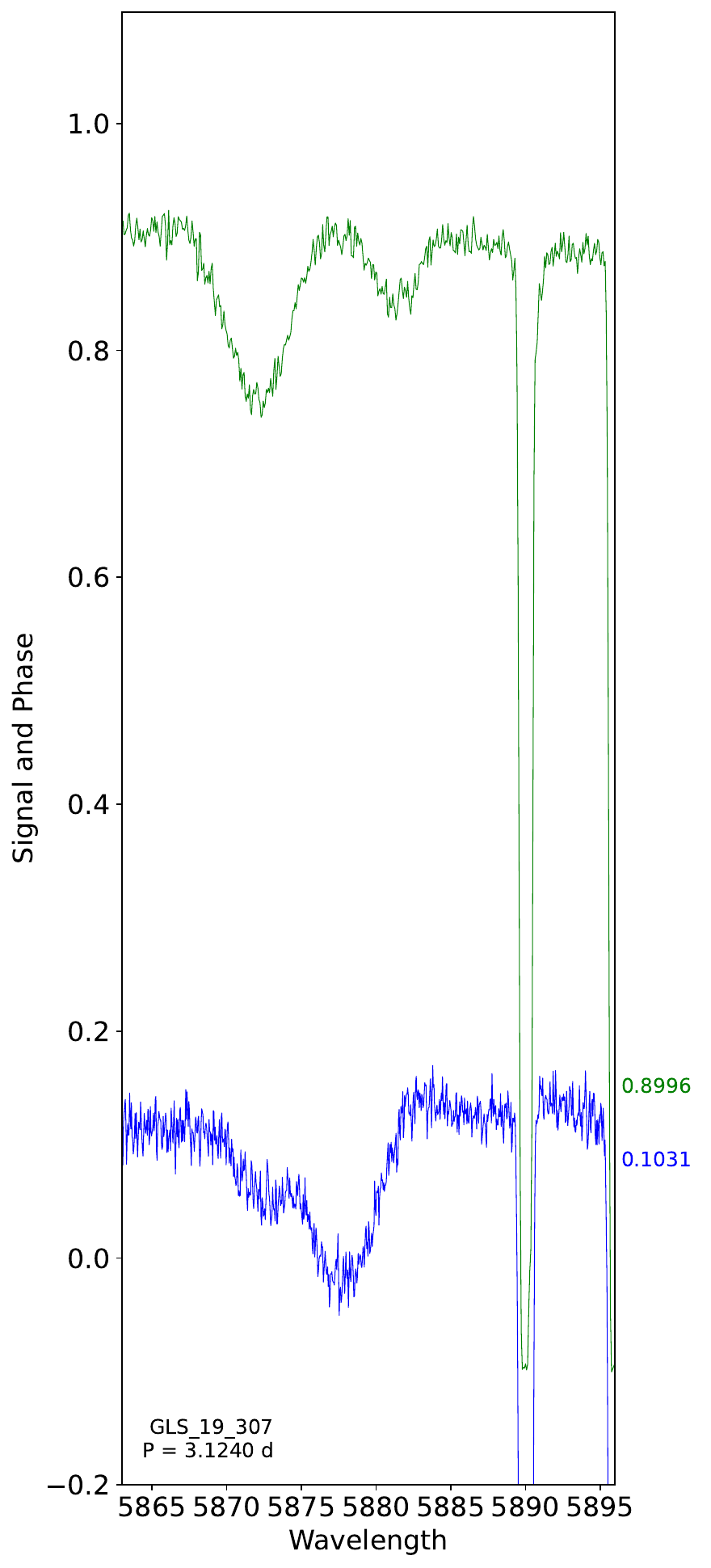}
    \caption{Same as Fig.~\ref{BD+61487_4541} for GLS~\num{19307} and for the \ion{He}{I} $\lambda$~5875 line. With only two spectra, it is already clear that this is an SB2 system.}
    \label{GLS19307_5876.pdf}
\end{figure}


\begin{figure}
    \centering
    \includegraphics[width=0.5\textwidth]{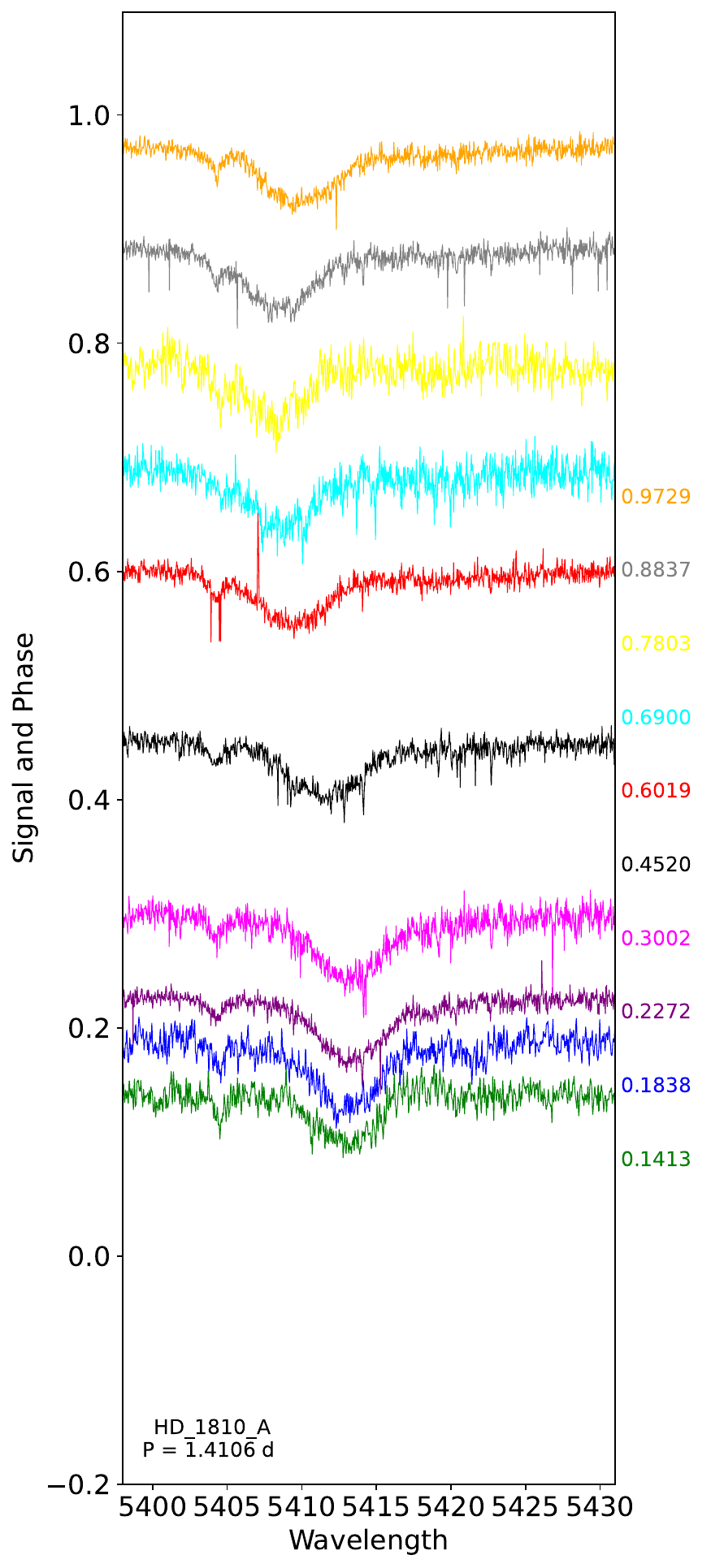}
    \caption{Same as Fig.~\ref{BD+61487_4541} for HD~\num[detect-all]{1810}~A and for the \ion{He}{II} $\lambda$~5411 line. This line is visible only in O-type stars.
    }
    \label{HD1810_5411}
\end{figure}


\begin{figure}
    \centering
    \includegraphics[width=0.5\textwidth]{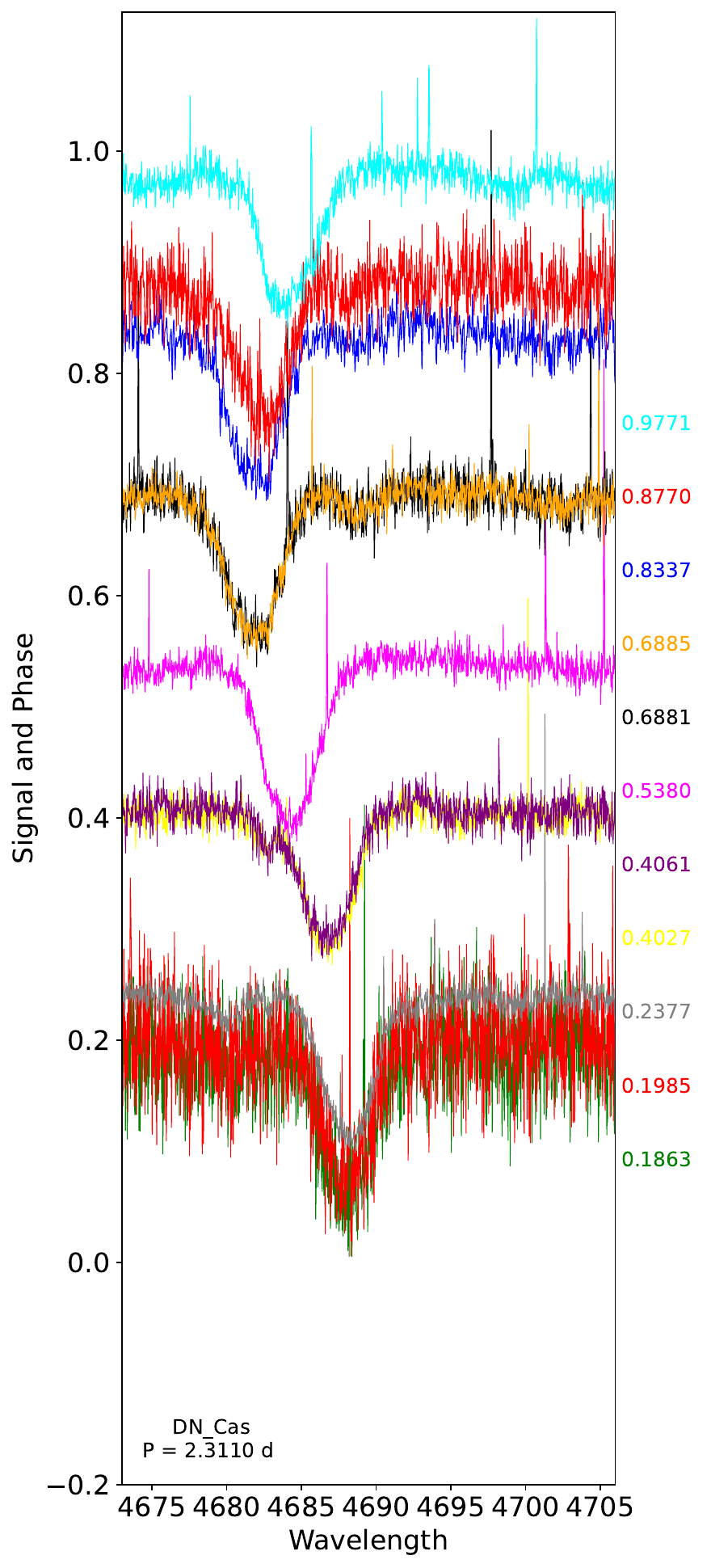}
    \caption{Same as Fig.~\ref{BD+61487_4541} for DN~Cas and for the \ion{He}{II} $\lambda$~4686 line. The companion is barely visible, indicating a B0 type.}
    \label{BD+60470_4686}
\end{figure}


\begin{figure*}
    \centering
    \includegraphics[width=\textwidth]{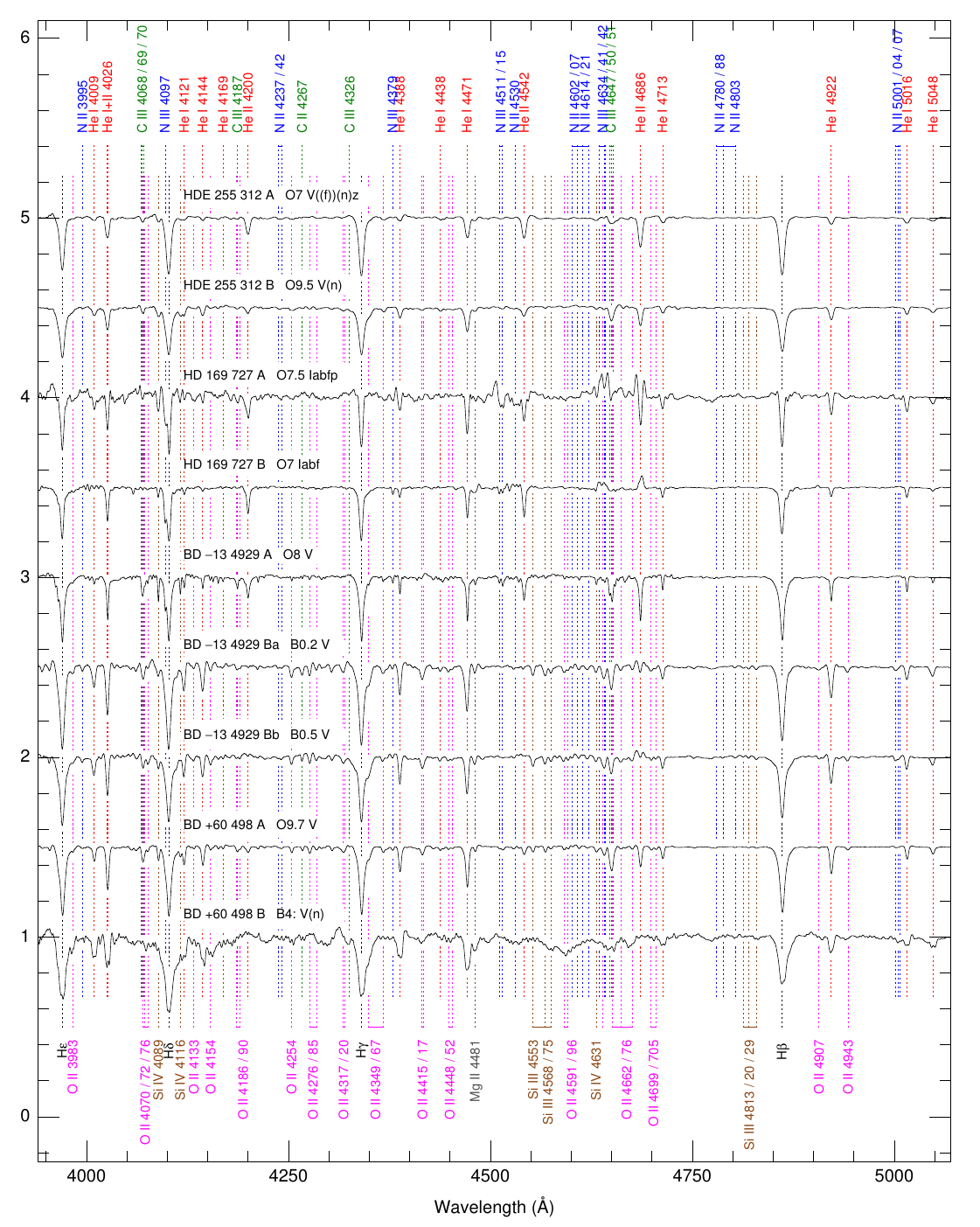}
    \caption{Disentangled spectra degraded to $R\sim 2500$ for spectral classification. As \textsc{unwind} has been used to extract them, the ISM lines have been eliminated.}
    \label{spectra1}
\end{figure*}

\addtocounter{figure}{-1}

\begin{figure*}
    \centering
    \includegraphics[width=\textwidth]{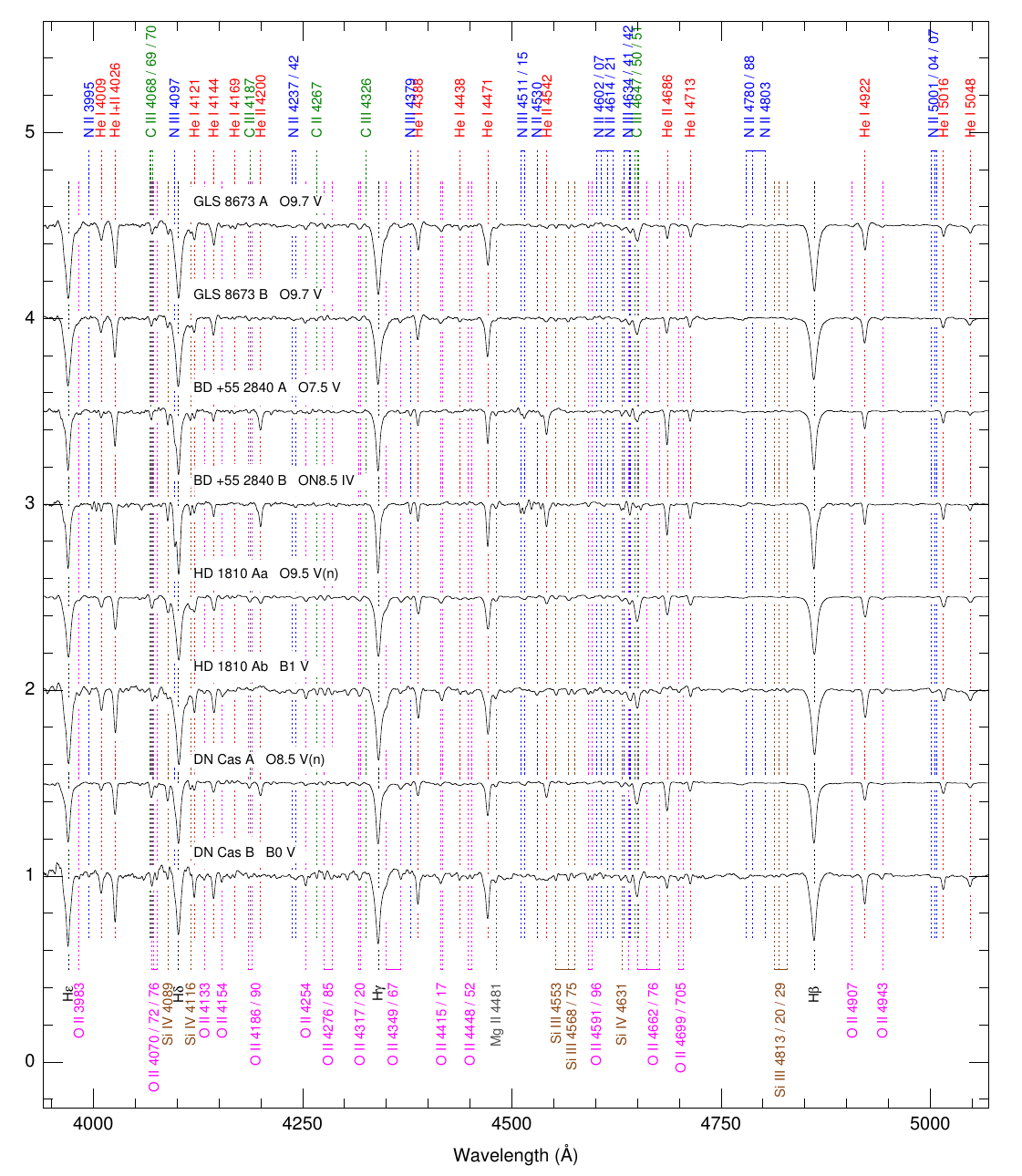}
    \caption{(Continued).}
\end{figure*}

\begin{figure*}
    \centering
    \includegraphics[width=\textwidth]{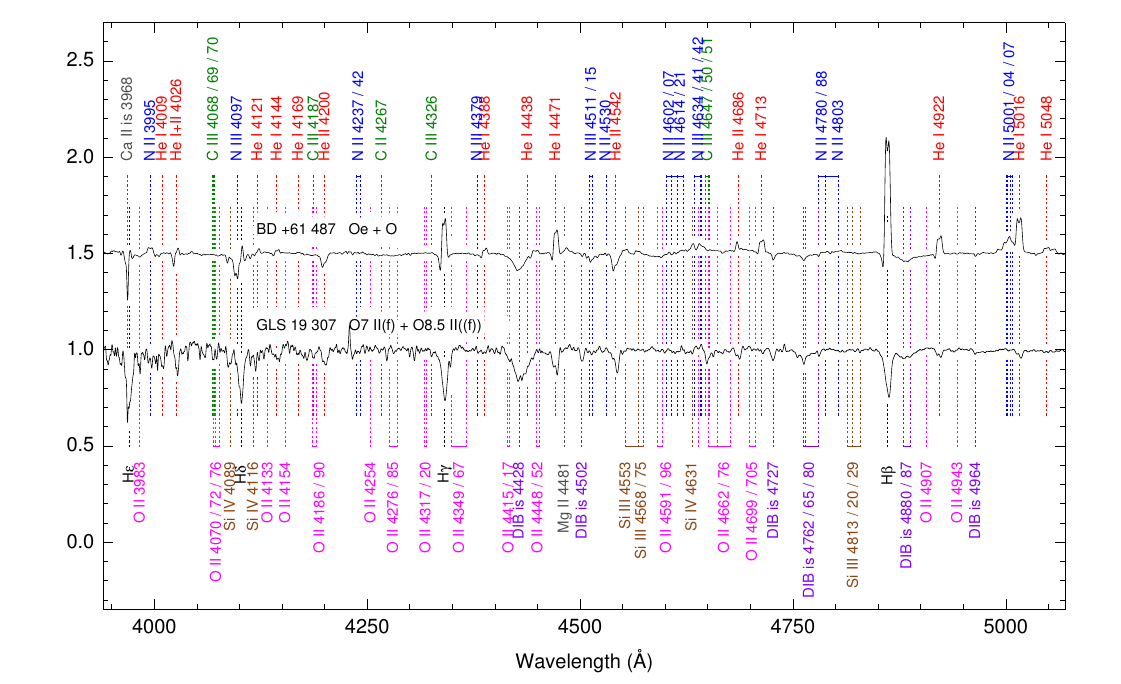}
    \caption{Non-disentangled spectra degraded to $R\sim 2500$ for spectral classification. ISM lines are present and indicated.}
    \label{spectra3}
\end{figure*}


\begin{figure*}
    \centering
    \includegraphics[width=\textwidth]{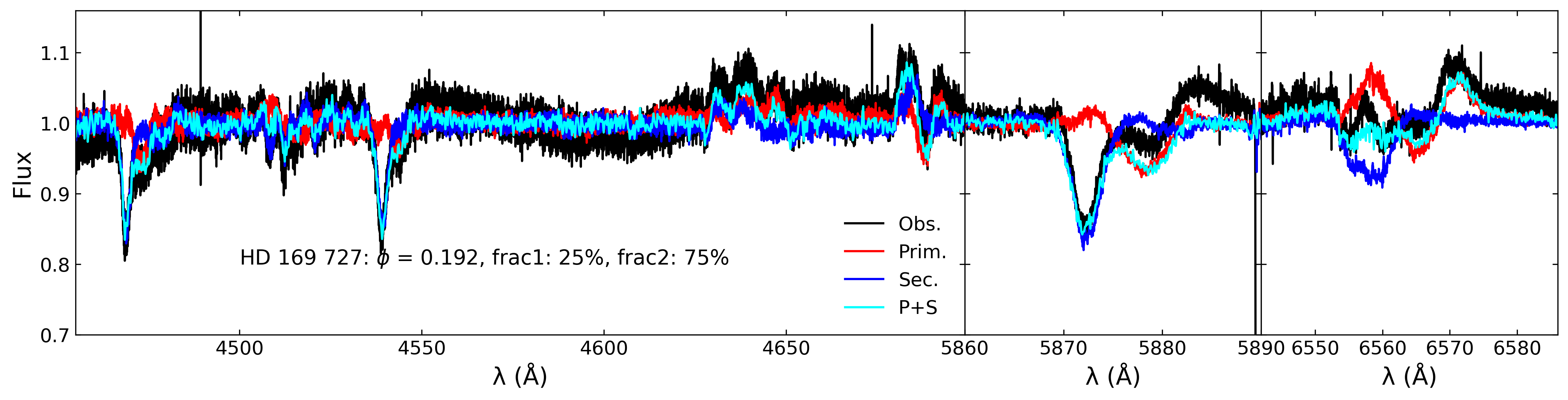}
    \includegraphics[width=\textwidth]{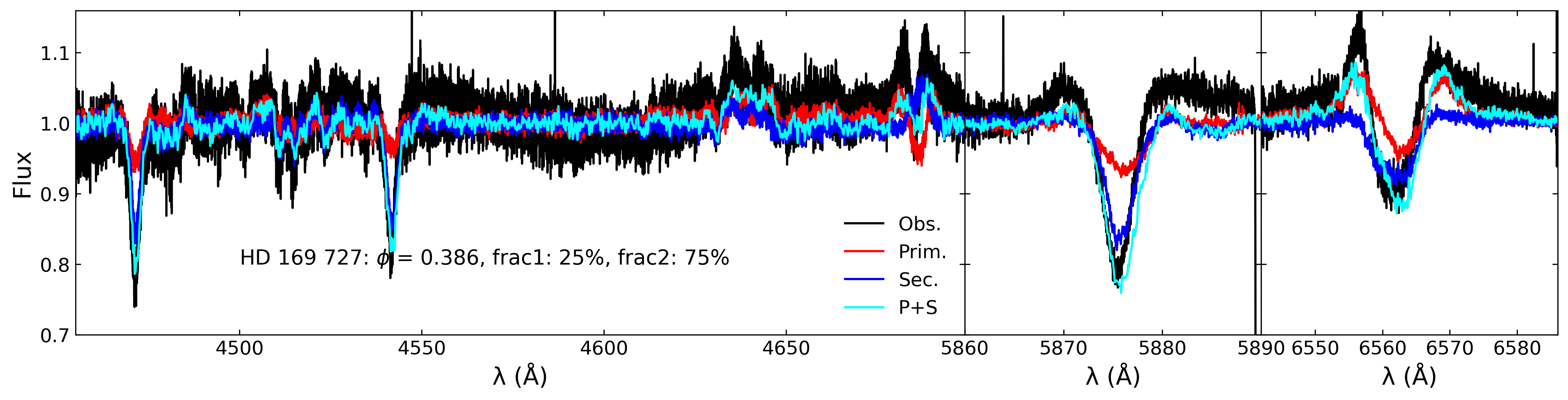}
    \caption{Same as Fig.~\ref{HDE255312_1} for HD~\num[detect-all]{169727}}.
    \label{HD169727_1}
\end{figure*}
\begin{figure*}
    \centering
    \includegraphics[width=\textwidth]{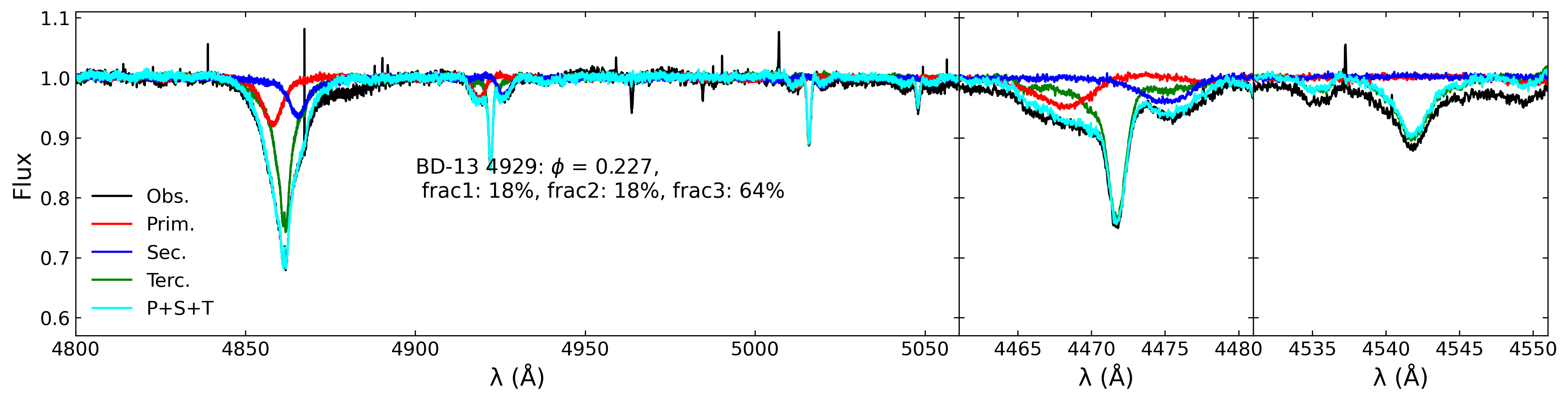}
    \includegraphics[width=\textwidth]{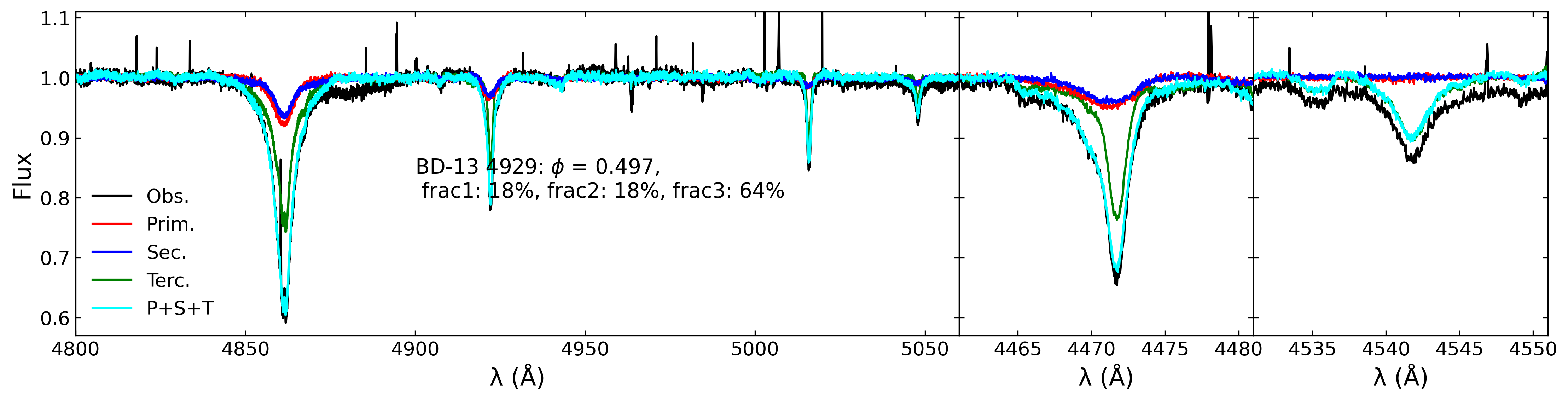}
    \caption{Same as Fig.~\ref{HDE255312_1} for BD~$-$13~4929.}
    \label{BD-134929_1s}
\end{figure*}
\begin{figure*}
    \centering
    \includegraphics[width=\textwidth]{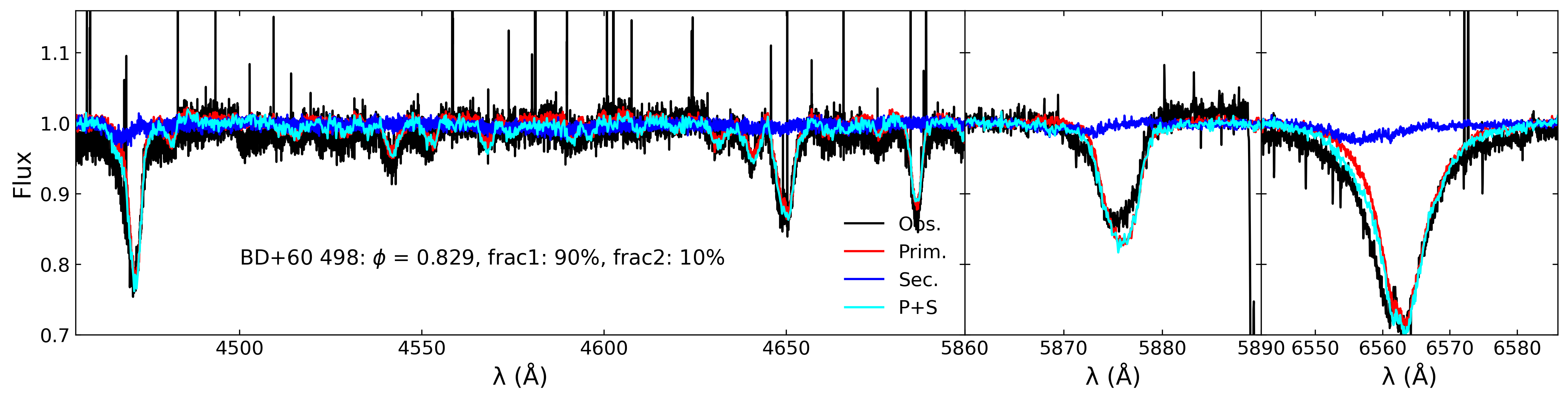}
    \includegraphics[width=\textwidth]{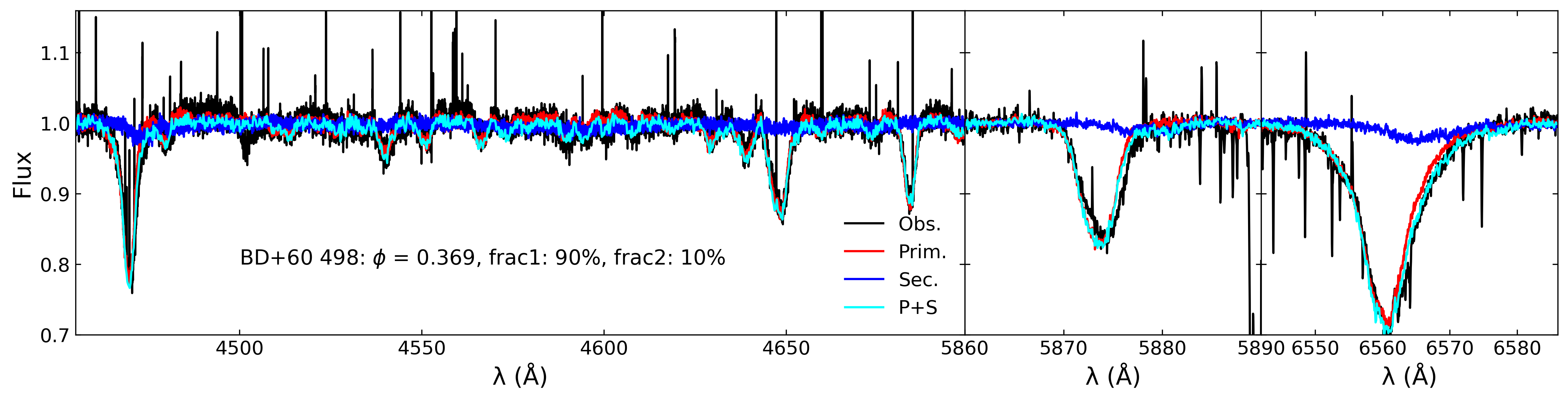}
    \caption{Same as Fig.~\ref{HDE255312_1} for BD~$+$60~498.}
    \label{BD+60498_1}
\end{figure*}
\begin{figure*}
    \centering
    \includegraphics[width=\textwidth]{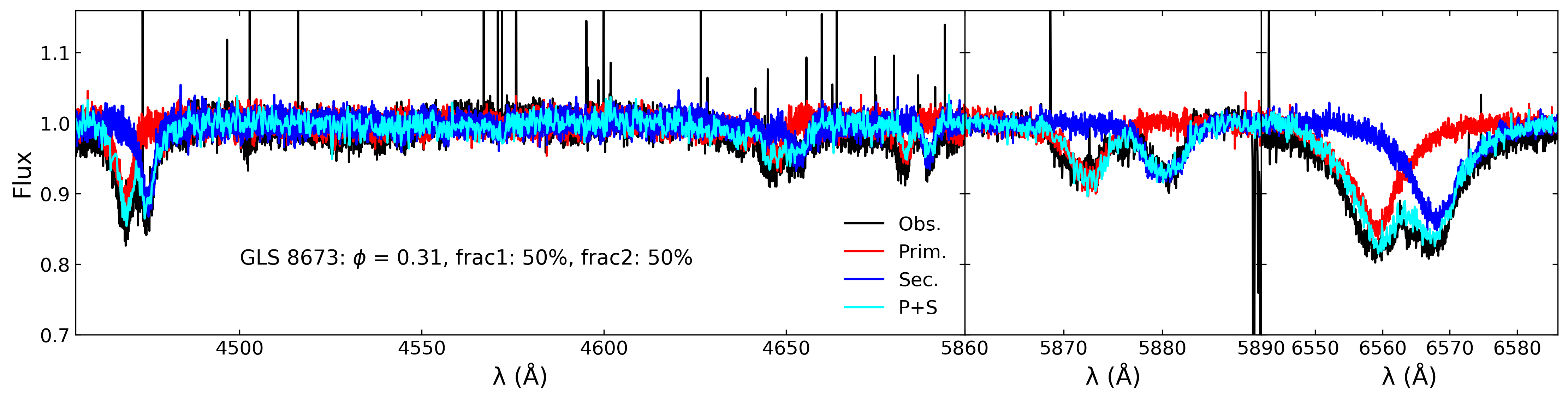}
    \includegraphics[width=\textwidth]{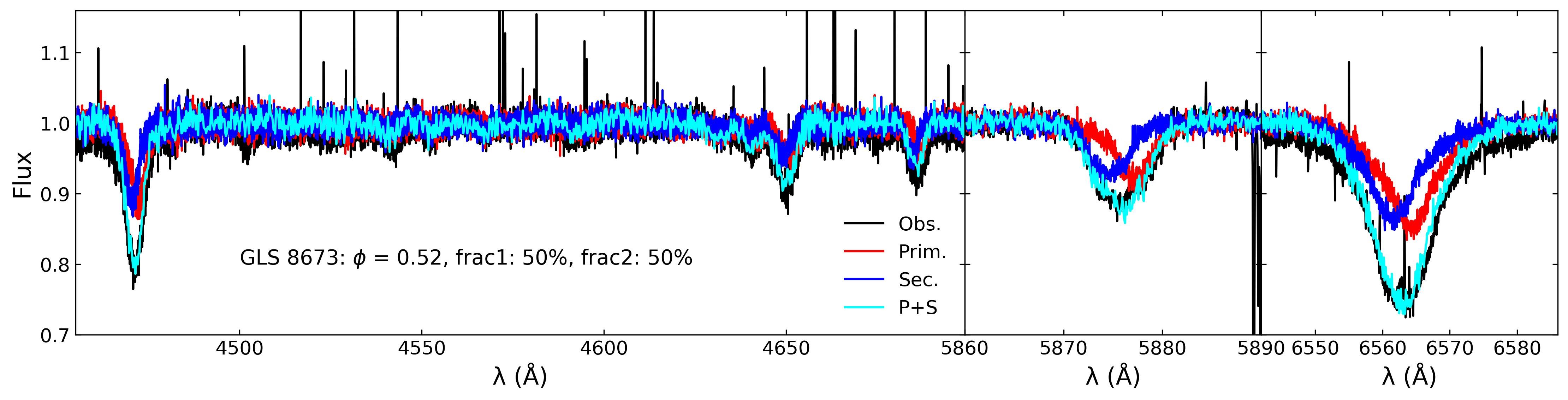}
    \caption{Same as Fig.~\ref{HDE255312_1} for GLS~8673.}
    \label{ALS8673_1}
\end{figure*}
\begin{figure*}
    \centering
    \includegraphics[width=\textwidth]{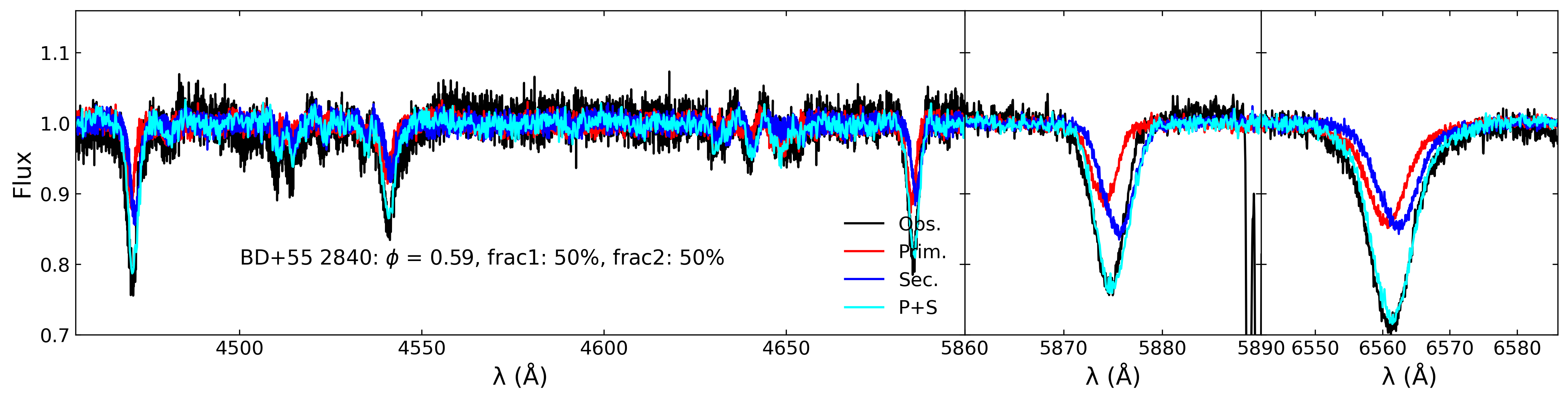}
    \includegraphics[width=\textwidth]{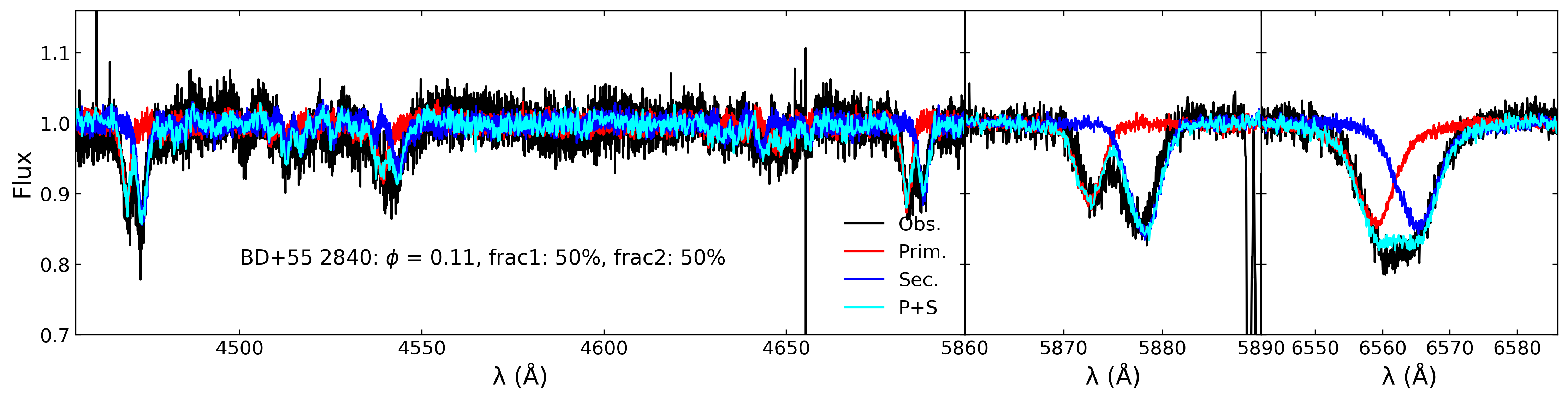}
    \caption{Same as Fig.~\ref{HDE255312_1} for BD~$+$55~2840.}
    \label{BD+552850_1}
\end{figure*}
\begin{figure*}
    \centering
    \includegraphics[width=\textwidth]{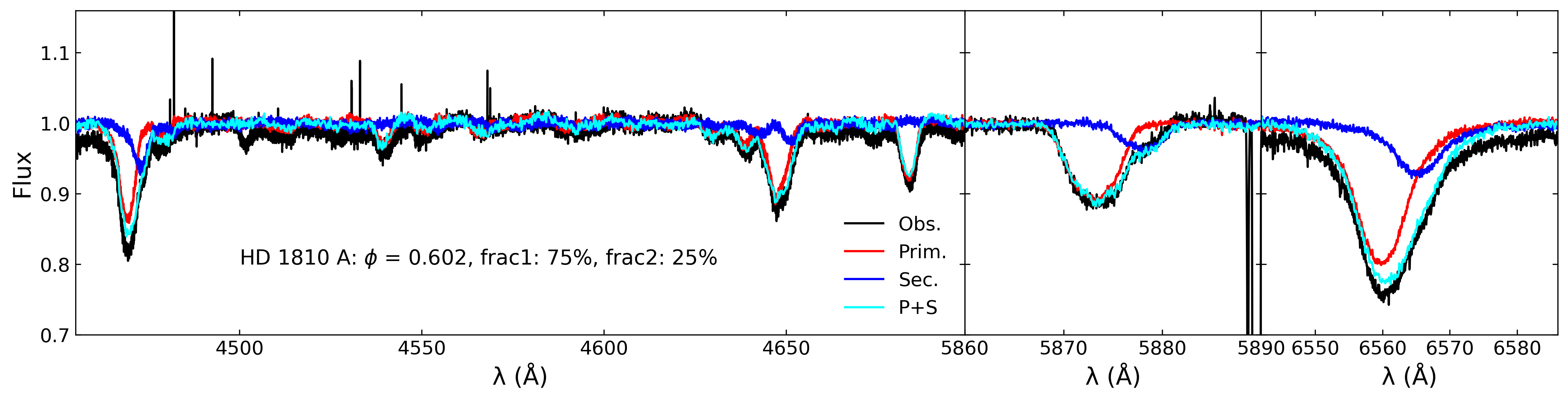}
    \includegraphics[width=\textwidth]{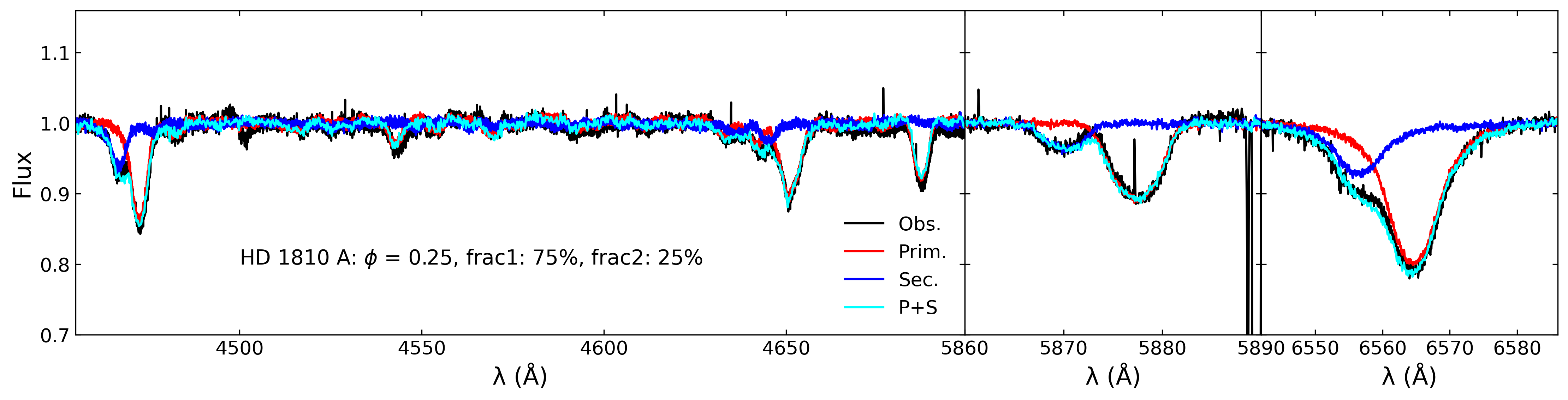}
    \caption{Same as Fig.~\ref{HDE255312_1} for HD~\num[detect-all]{1810}~A.}
    \label{HD1810A_1}
\end{figure*}
\begin{figure*}
    \centering
    \includegraphics[width=\textwidth]{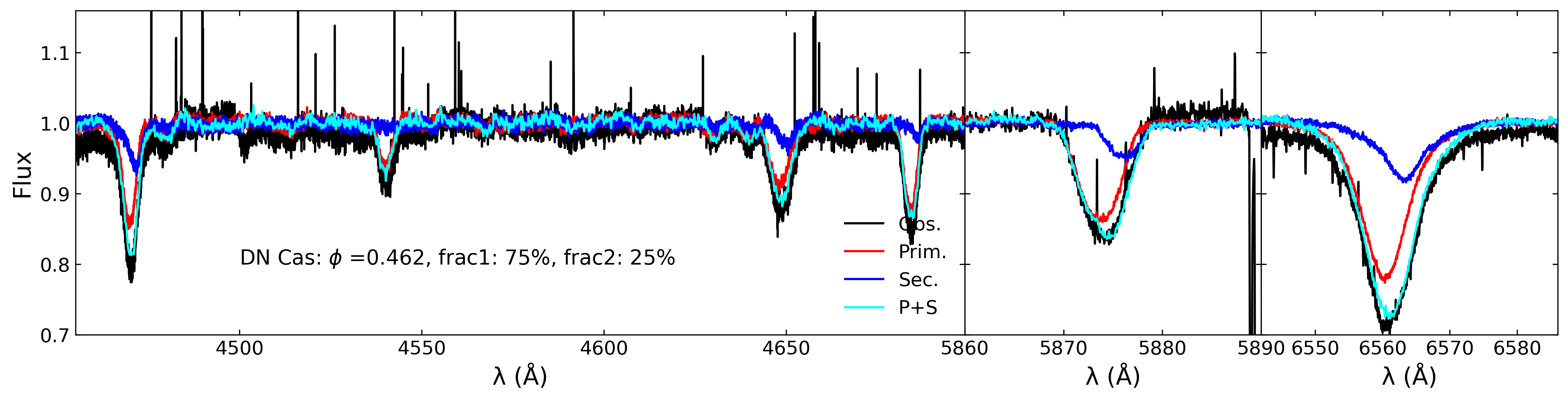}
    \includegraphics[width=\textwidth]{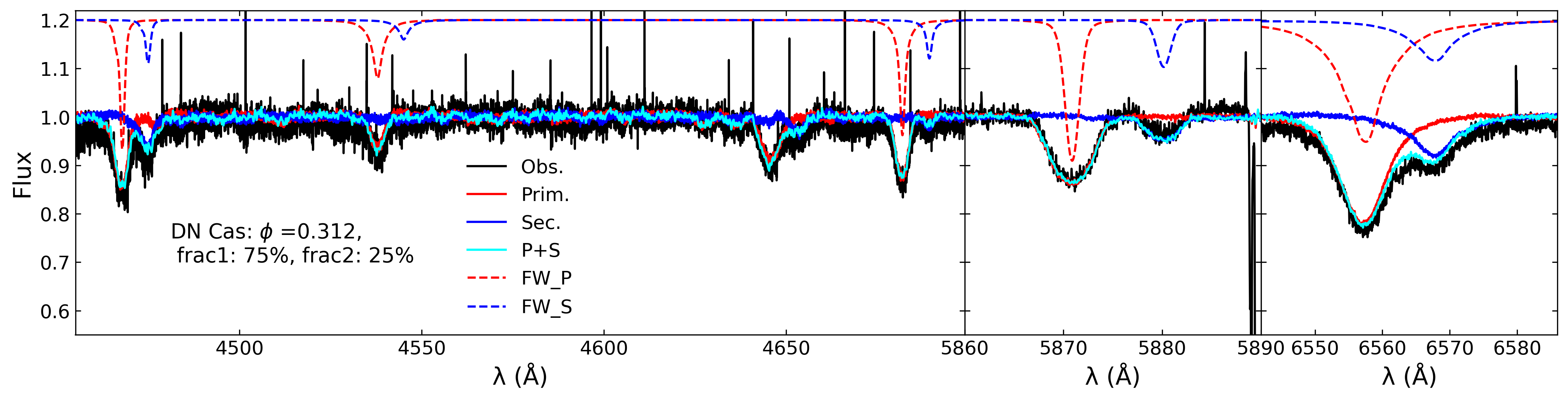}
    \caption{Same as Fig.~\ref{HDE255312_1} for DN~Cas. This plot also includes synthetic atmosphere models computed with {\sc{fastwind}}, employed to confirm via cross-correlation analysis that the spectral line shifts in the disentangled spectra are in excellent agreement with those of the models. The {\sc{fastwind}} model is shifted for clarity, and computed with low rotational broadening to highlight the core position of the lines. The model includes only H and He lines.}
    \label{DN_Cas_1}
\end{figure*}

\end{appendix}

\end{document}